\documentclass[12pt]{article}

\usepackage{amsmath}
\usepackage{amssymb}
\usepackage{booktabs}
\usepackage{graphicx}
\usepackage{multirow}
\usepackage{threeparttable}
\usepackage{rotating}
\usepackage[authoryear]{natbib}
\usepackage{geometry}
\usepackage{setspace}
\usepackage[hidelinks]{hyperref}
\hypersetup{pdfauthor={Taojun Hu}, pdftitle={Improving the selection-function-based sensitivity analysis for publication bias with restricted maximum likelihood}}
\newcommand{\MLEunadj}{\mathrm{MLE\text{-}unadj}}
\newcommand{\REMLunadj}{\mathrm{REML\text{-}unadj}}
\newcommand{\MLEadj}{\mathrm{MLE\text{-}adj}}
\newcommand{\REMLadj}{\mathrm{REML\text{-}adj}}
\newcommand{\pkg}[1]{{\normalfont\ttfamily #1}}

\title{\bf Improving the selection-function-based sensitivity analysis for publication bias with restricted maximum likelihood}

\author{Taojun Hu$^{1,2}$, Yi Zhou$^{3,2}$, Xiao-Hua Zhou$^{4,5,*}$, and Satoshi Hattori$^{2,6,*}$ \\[6pt]
\parbox{0.95\textwidth}{\centering\small\setstretch{1.3}
$^{1}$School of Medicine, Chongqing University, 400030, Chongqing, China\\
$^{2}$Department of Biomedical Statistics, Graduate School of Medicine, The University of Osaka, Yamadaoka 2-2, 565-0871, Osaka, Japan\\
$^{3}$Division of Mathematics and Informatics, Graduate School of Human Development and Environment, Kobe University, Rokkodai-cho 1-1, 657-8501, Kobe, Japan\\
$^{4}$Beijing International Center for Mathematical Research, Peking University, 100871, Beijing, China\\
$^{5}$Department of Biostatistics, School of Public Health, Peking University, 100191, Beijing, China\\
$^{6}$Integrated Frontier Research for Medical Science Division, Institute for Open and Transdisciplinary Research Initiatives (OTRI), The University of Osaka, Yamadaoka 2-2, 565-0871, Osaka, Japan\\
$^{*}$Correspondence: Xiao-Hua Zhou, azhou@math.pku.edu.cn; Satoshi Hattori, hattoris@biostat.med.osaka-u.ac.jp}}

\date{}

\begin{document}

\maketitle

\begin{abstract}
In random-effects meta-analysis, the between-study variance quantifies the heterogeneity among studies. Without selective publication, the restricted maximum likelihood (REML) method is recommended for this variance, since the maximum likelihood (ML) method underestimates it when the number of studies is small. However, under selective publication, referred to as publication bias (PB), the published studies no longer follow the normal-normal model, and the REML cannot be directly applied. Thus, the existing sensitivity analysis methods for PB based on selection functions estimate the between-study variance by the ML conditional on published studies and inherit its underestimation. This underestimation also affects the estimate of the common effect and its confidence interval. To improve the estimation of the common effect and the heterogeneity, we propose a REML-type estimator of the between-study variance for the selection-function-based sensitivity analysis. It centers the profile score by its approximate mean and reduces exactly to the REML without PB. The proposed sensitivity analysis method is a more accurate alternative to the ML-based sensitivity analysis in meta-analyses with a small number of studies, and it is the first REML-type estimator in the presence of PB. Simulation studies substantiate that the proposed method reduces the median bias of the between-study variance and improves the estimation error and coverage probability for the common effect compared with the conditional ML. We illustrate the proposed method with two meta-analyses on high maintenance-dose clopidogrel and nicotine gum for smoking cessation. The proposed method would give useful guidance on the between-study heterogeneity in sensitivity analysis for PB.

\end{abstract}

\section{Introduction}
\label{sec.1}

Meta-analysis is a technique to integrate the results of multiple studies and establish evidence of higher precision. It is widely applied in medicine, sociology, and psychology~\citep{dersimonian2015meta,mazei2015meta,lipsey2017efficacy}. The random-effects model~\citep{hedges1983random} has become the dominant approach in meta-analysis~\citep{schmidt2009fixed,clark2015should}. It has a two-level structure: the within-study model describes the random error of each study, and the between-study model describes the variation of the study-specific effects around the common effect size. The between-study variance $\tau^2$ quantifies this heterogeneity. It is of interest in meta-analysis since it indicates whether the effect is consistent across populations and quantifies the population heterogeneity. It also determines the weights of the studies, and thus the estimate of the common effect size and the width of its confidence interval (CI).

Numerous estimators of $\tau^2$ have been proposed~\citep{dersimonian1986meta,dersimonian2007random,hardy1996likelihood,knapp2006assessing,viechtbauer2007confidence}, together with derived measures such as the $I^2$ statistic~\citep{higgins2002quantifying,huedo2006assessing}. The maximum likelihood (ML) and restricted maximum likelihood (REML) estimators are representative methods~\citep{hardy1996likelihood,viechtbauer2005bias}. The ML estimator has been proved to underestimate $\tau^2$ in finite samples~\citep{viechtbauer2005bias}. The REML removes this bias by maximizing the likelihood of error contrasts, which is the linear combination of the observed effect sizes. Since most meta-analyses include only a few studies~\citep{davey2011characteristics}, this difference is not negligible in practice. The REML estimator has shown good performance in simulation studies~\citep{langan2015empirical,langan2019comparison}, is widely recommended, and is the default estimator in the \pkg{R} package \pkg{metafor}~\citep{viechtbauer2010conducting}.

Publication bias (PB), the bias caused by selective publication of studies with statistically significant results or otherwise, is one of the most serious threats to meta-analysis~\citep{begg1988publication}. A number of methods have been proposed to detect or adjust for PB, including the trim-and-fill method~\citep{duval2000trim}, Egger's regression test~\citep{egger1997bias}, and Begg's rank correlation test~\citep{begg1988publication}. Many quantitative approaches rely on selection functions that describe the probability of each study being published~\citep{jin2015statistical}. Among them, sensitivity analysis methods assess how the integrated results change with different levels of pre-specified sensitivity parameters, like the number of possibly unpublished studies~\citep{copas2004bound,copas2013likelihood}. \citet{copas2013likelihood} proposed a likelihood-based sensitivity analysis with a $t$-statistic-based selection function, in which the parameters are estimated by maximizing the likelihood conditional on published studies for a given marginal selection probability. This sensitivity analysis method based on the $t$-statistic selection function has also been extended to meta-analyses of diagnostic test accuracy and meta-analyses of sparse events~\citep{zhou2023likelihood,hu2024likelihood,hu2024sensitivity}.

However, these methods focus on the common effect size and pay little attention to the heterogeneity parameter. Besides, all these methods estimate the parameters, including $\tau^2$, by the ML conditional on published studies. Thus, they are expected to inherit the drawback of the ML in the standard setting, that is, the underestimation of the heterogeneity. In addition, $\tau^2$ determines the weights of studies in estimating the common effect size; thus, the underestimation of $\tau^2$ also affects the estimate of the common effect size and its CI. \citet{honekopp2022heterogeneity} evaluated the robustness of several heterogeneity estimators not designed for PB; the REML was relatively robust and was recommended, but it still showed substantial bias under PB. This suggests the potential of introducing the REML into methods addressing PB, such as the $t$-statistic-based sensitivity analysis of~\citet{copas2013likelihood}, to further reduce the bias. A natural choice would be the REML under the selection model. However, the REML cannot be directly applied: since the publication probability depends on the $t$-statistic of each study, the published effect sizes do not follow the normal-normal (NN) model, and error contrasts whose distribution should be free of the common effect size are not available. To the best of our knowledge, there is no REML-type estimator of the heterogeneity parameter available for the selection-function-based sensitivity analysis for PB.

In this work, we propose the first REML-type estimator of $\tau^2$ for the sensitivity analysis of~\citet{copas2013likelihood}, based on the adjusted profile likelihood of~\citet{mccullagh1990simple}. Instead of following the convention of the conditional ML, the proposed method brings the REML, the recommended estimator in the primary analysis, into the sensitivity analysis for PB. It is particularly useful for meta-analyses with a small number of studies, which are the majority in practice and in which the underestimation by the ML is the most serious. The key observation is that the REML estimating equation is the ML score for $\tau^2$, with the common effect size replaced by its estimate, centered by its mean. Following this, we center the profile score for $\tau^2$ under the conditional likelihood by its approximate mean. The adjustment accounts for the estimation of both the common effect size and the selection parameter under PB. In the absence of PB, the proposed estimator reduces exactly to the REML; thus, the sensitivity analysis starts from the REML estimate reported in the primary analysis. In addition, we add a Jeffreys-type penalty~\citep{firth1993bias,kosmidis2021jeffreys} to prevent the estimate of the selection parameter from diverging, which frequently occurs with the MLE in small meta-analyses. The rest of this article is organized as follows. In Section~\ref{sec.2}, we review the random-effects model, the ML and REML estimators, and the sensitivity analysis method of~\citet{copas2013likelihood}, which estimates the parameters by the ML. In Section~\ref{sec.3}, we develop the proposed heterogeneity estimator. In Section~\ref{sec.4}, we conduct simulation studies to compare the proposed estimator with the existing estimators. In Section~\ref{sec.5}, we illustrate the proposed method with two real meta-analyses. We end with discussions in Section~\ref{sec.6}.

\section{Random-effects models and the sensitivity analysis for publication bias}\label{sec.2}

\subsection{The conventional random-effects model and the heterogeneity estimators}\label{sec.2.1}

The conventional random-effects NN model assumes that both the within-study and between-study models follow normal distributions. Assuming there are $N$ published studies, the observed effect size $\hat{\theta}_i$ ($i = 1, \cdots, N$) can be decomposed into two parts: the within-study model $\hat{\theta}_i \sim N(\theta_i, \sigma_i^2)$, where $\theta_i$ is the true study-specific effect size in the $i$-th study and $\sigma_i^2$ is the within-study variance; and the between-study model $\theta_i\sim N(\theta, \tau^2)$, where $\theta$ is the common effect size and $\tau^2$ is the between-study variance. The marginal model is $\hat{\theta}_i \sim N(\theta, \sigma_i^2+\tau^2)$. We follow the convention in meta-analyses: $\sigma_i^2$ is regarded as known although it is estimated in each study.

Let $w_i=1/(\sigma_i^2+\tau^2)$ and $W=\sum_{i=1}^N w_i$. For a given $\tau^2$, the common effect size is estimated by the inverse-variance weighted mean
\begin{equation}
    \hat{\theta} = \frac{\sum_{i=1}^N w_i \hat{\theta}_i}{W}.
    \label{eq.1}
\end{equation}
The ML estimator maximizes the marginal log-likelihood $\ell_{\mathrm{NN}}(\theta, \tau^2)$. Its score for $\tau^2$ is
\begin{equation}
    \frac{\partial \ell_{\mathrm{NN}}}{\partial \tau^2}=\frac{1}{2}\sum_{i=1}^N \left\{w_i^2\left(\hat{\theta}_i-\theta\right)^2-w_i\right\},
    \label{eq.2}
\end{equation}
and setting it to zero with $\theta$ replaced by $\hat{\theta}$ in~\eqref{eq.1} gives
\begin{equation}
\hat{\tau}^2_{\MLEunadj}=\frac{\sum_{i=1}^N w_i^2\left\{\left(\hat{\theta}_i-\hat{\theta}\right)^2-\sigma_i^2\right\}}{\sum_{i=1}^N w_i^2}.
    \label{eq.3}
\end{equation}
The ML estimator iterates between~\eqref{eq.1} and \eqref{eq.3} until convergence. Using the estimated $\hat{\theta}$ in place of the true $\theta$ leads to a downward bias of $\hat{\tau}^2_{\MLEunadj}$ in finite samples. The REML estimator addresses this by maximizing the likelihood of error contrasts~\citep{patterson1971recovery,harville1974bayesian}. An error contrast is a linear combination of the observed effect sizes,
\begin{equation}
    \sum_{i=1}^N a_i\hat{\theta}_i\quad\text{with}\quad \sum_{i=1}^N a_i=0.
    \label{eq.4}
\end{equation}
Under the NN model, the error contrast~\eqref{eq.4} follows $N\{0,\sum_{i=1}^N a_i^2(\sigma_i^2+\tau^2)\}$, whose distribution depends on $\tau^2$ but not on $\theta$. The REML estimator of $\tau^2$ maximizes the likelihood of $N-1$ linearly independent error contrasts, and thus it does not involve the estimation of $\theta$~\citep{viechtbauer2005bias}. It yields the modified form
\begin{equation}
\hat{\tau}^2_{\REMLunadj}=\frac{\sum_{i=1}^N w_i^2\left\{\left(\hat{\theta}_i-\hat{\theta}\right)^2-\sigma_i^2\right\}}{\sum_{i=1}^N w_i^2}+\frac{1}{W} .
\label{eq.5}
\end{equation}
Equivalently, $\hat{\tau}^2_{\REMLunadj}$ solves
\begin{equation}
    \frac{\partial \ell_{\mathrm{NN}}}{\partial \tau^2}\bigg|_{\theta=\hat{\theta}}+\frac{\sum_{i=1}^N w_i^2}{2W}=0,
    \label{eq.6}
\end{equation}
where the added term is minus the expectation of~\eqref{eq.2} at $\theta=\hat{\theta}$. That is, the REML centers the score for $\tau^2$ by its mean after plugging in $\hat{\theta}$, whereas the ML does not. Both the ML and REML estimators truncate negative estimates of $\tau^2$ to 0. Since they do not account for PB, we refer to them as the unadjusted ML and REML estimators, and denote the corresponding estimates of $\theta$, given by~\eqref{eq.1} with $\tau^2$ replaced by $\hat{\tau}^2_{\MLEunadj}$ and $\hat{\tau}^2_{\REMLunadj}$, by $\hat{\theta}_{\MLEunadj}$ and $\hat{\theta}_{\REMLunadj}$, respectively.

\subsection{The likelihood-based sensitivity analysis method for publication bias}\label{sec.2.2}

\citet{copas2013likelihood} proposed a $t$-statistic-based sensitivity analysis to assess how selective publication in random-effects meta-analyses affects inference. They assume the publication process is determined by the reported $p$-value or the $t$-statistic of a study. As studies with more significant results are more likely published, this assumption is reasonable and the resulting sensitivity analysis is interpretable. They used the selection function
\begin{equation}
    \mathrm{Pr}\left(\operatorname{select} \mid \hat{\theta}_i, \sigma_i\right)=\Phi\left(\alpha+\beta t_i\right),
    \label{eq.7}
\end{equation}
where $t_i=\hat{\theta}_i/\sigma_i$ is the $t$-statistic of the $i$-th study and $\Phi(\cdot)$ is the cumulative distribution function of the standard normal distribution. Thus, the publication probability is monotone in the $t$-statistic or the corresponding $p$-value. With this two-parameter probit selection function, the marginal selection probability given only $\sigma_i$ has the explicit form
\begin{equation}
    \mathrm{Pr}\left(\operatorname{select} \mid \sigma_i\right)=\Phi\left(\frac{\gamma_{1i}}{\gamma_{2i}}\right),\quad \gamma_{1i} = \alpha + \beta\theta/\sigma_i, \quad \gamma_{2i} = \sqrt{ 1 + \beta^2(1 + \tau^2/\sigma_i^2) }.
    \label{eq.8}
\end{equation}
Their sensitivity analysis relies on the marginal selection probability $p=\mathrm{Pr}(\mathrm{select})$ as the sensitivity parameter. The parameter $p$ can be interpreted as the ratio of published studies among all published and unpublished studies. Thus, one can assess how many unpublished studies would be required to overturn the conclusions. Given $p$, \citet{copas2013likelihood} showed that $1/p = \boldsymbol{E}_{o}\{1/\mathrm{Pr}(\mathrm{select} \mid \sigma_i)\}$, where $\boldsymbol{E}_o$ is the expectation over published studies, and used its empirical version as a constraint:
\begin{equation}
    \frac{1}{p} = \frac{1}{N} \sum_{i=1}^N \frac{1}{\Phi\left(\gamma_{1i}/\gamma_{2i}\right)},
    \label{eq.9}
\end{equation}
where $N$ now denotes the number of published studies. Since only published studies are available, the likelihood conditional on published studies is used. Omitting the terms free of the parameters, the log-likelihood conditional on published studies is
\begin{equation}
    \ell_o(\theta, \tau^2, \alpha, \beta) = \sum_{i=1}^N \left\{ - \frac{1}{2} \log (\sigma_i^2+\tau^2) - \frac{1}{2} \frac{(\hat{\theta}_i - \theta)^2}{\sigma_i^2 + \tau^2} + \log \Phi\left(\alpha + \beta \frac{\hat{\theta}_i}{\sigma_i} \right) - \log \Phi\left(\frac{\gamma_{1i}}{\gamma_{2i}}\right) \right\}.
\label{eq.10}
\end{equation}
\citet{copas2013likelihood} estimated $(\theta, \tau^2, \alpha, \beta)$ by maximizing~\eqref{eq.10} subject to the constraint~\eqref{eq.9}. Since the right-hand side of~\eqref{eq.9} is monotone in $\alpha$, the constraint can be solved for $\alpha$ given the other three parameters; we denote the solution by $\hat{\alpha}(\theta, \tau^2, \beta)$. Substituting it into~\eqref{eq.10}, we obtain the conditional log-likelihood of $\xi=(\xi_1,\xi_2,\xi_3)=(\theta, \tau^2, \beta)$,
\begin{equation}
    \ell_{\mathrm{cond}}(\theta, \tau^2, \beta)=\ell_o\{\theta, \tau^2, \hat{\alpha}(\theta, \tau^2, \beta), \beta\},
    \label{eq.11}
\end{equation}
which incorporates the constraint~\eqref{eq.9}. The estimator of~\citet{copas2013likelihood} maximizes $\ell_{\mathrm{cond}}$ over $\xi$. We refer to it as the MLE hereafter and denote its estimates of $\theta$ and $\tau^2$ by $\hat{\theta}_{\MLEadj}$ and $\hat{\tau}^2_{\MLEadj}$; in the absence of PB, they reduce to $\hat{\theta}_{\MLEunadj}$ and $\hat{\tau}^2_{\MLEunadj}$ in Section~\ref{sec.2.1}.

\section{A REML-type estimator of the between-study variance under publication bias}\label{sec.3}

\subsection{Overview}\label{sec.3.1}

Our estimator is defined with the conditional log-likelihood $\ell_{\mathrm{cond}}$ in~\eqref{eq.11} with the MLE. We denote its score functions by
\begin{equation}
    U_\theta(\xi)=\frac{\partial \ell_{\mathrm{cond}}(\xi)}{\partial\theta},\qquad U_{\tau^2}(\xi)=\frac{\partial \ell_{\mathrm{cond}}(\xi)}{\partial\tau^2},\qquad U_\beta(\xi)=\frac{\partial \ell_{\mathrm{cond}}(\xi)}{\partial\beta},
    \label{eq.12}
\end{equation}
whose explicit forms are given in Web Appendix A. Note that all three scores are derivatives of $\ell_{\mathrm{cond}}$, and thus they include the derivatives of $\hat{\alpha}(\theta, \tau^2, \beta)$. Given $\sigma_i$, each score has mean zero at the true $\xi$.

We regard $\tau^2$ as the parameter of interest, and let $\lambda=(\lambda_1,\lambda_2)=(\theta,\beta)$ denote the remaining two parameters. For a given $\tau^2$, let $\hat{\theta}(\tau^2)$ and $\hat{\beta}(\tau^2)$ solve $U_\theta=0$ and $U_\beta=0$, that is, it is satisfied that $U_\theta(\hat{\theta}(\tau^2), \tau^2, \hat{\beta}(\tau^2))=0$, and let $\hat{\lambda}=\{\hat{\theta}(\tau^2),\hat{\beta}(\tau^2)\}$. We define the profile score for $\tau^2$ by
\begin{equation}
    U^{\mathrm{PL}}_{\tau^2}(\tau^2)=U_{\tau^2}\{\hat{\theta}(\tau^2),\tau^2,\hat{\beta}(\tau^2)\},
    \label{eq.13}
\end{equation}
which is the derivative of the profile log-likelihood (PL) of $\tau^2$. The MLE $\hat{\tau}^2_{\MLEadj}$ solves $U^{\mathrm{PL}}_{\tau^2}(\tau^2)=0$. However, the profile score is biased. While $\boldsymbol{E}\{U_{\tau^2}(\theta,\tau^2,\beta)\}=0$ at the true value of $(\theta,\tau^2,\beta)$, the profile score~\eqref{eq.13} loses this unbiasedness property, since $\theta$ and $\beta$ are replaced with $\hat{\theta}(\tau^2)$ and $\hat{\beta}(\tau^2)$, respectively. In other words,
\begin{equation}
    \boldsymbol{E}\{U^{\mathrm{PL}}_{\tau^2}(\tau^2)\}\ne 0.
    \label{eq.14}
\end{equation}
This mean is of at least order 1, whereas the information for $\tau^2$ is of order $N$ (Web Appendix A). Thus, $\hat{\tau}^2_{\MLEadj}$ has a bias of at least order $N^{-1}$, which is not negligible when only a few studies are published. Without PB, there is no $\beta$, $\hat{\theta}(\tau^2)$ is~\eqref{eq.1}, and the profile score is~\eqref{eq.2} at $\theta=\hat{\theta}$, whose mean is exactly $-\sum_{i=1}^N w_i^2/(2W)$. The REML equation~\eqref{eq.6} subtracts this mean from the profile score; that is, the REML solves the profile score equation after centering the profile score by its mean.

The main idea of our proposal follows this observation. Under PB, a published $\hat{\theta}_i$ follows the conditional distribution given selection, which is not normal and whose mean depends on $\theta$ and $\sigma_i$ nonlinearly through the selection function~\eqref{eq.7}. Thus, unlike under the NN model in Section~\ref{sec.2.1}, the distribution of an error contrast~\eqref{eq.4} depends on $\theta$, and the REML cannot be constructed from error contrasts. Instead, we follow the characterization of the REML in~\eqref{eq.6} and center the profile score~\eqref{eq.13} by its approximate mean, since the exact mean is hard to derive. We give the proposed estimating equation in Section~\ref{sec.3.2}, construct the approximate mean in Section~\ref{sec.3.3}, add a Jeffreys-type penalty for small numbers of studies in Section~\ref{sec.3.4}, and describe the computation and inference in Section~\ref{sec.3.5}.

\subsection{The proposed estimator}\label{sec.3.2}

Following the characterization of the REML~\eqref{eq.6} and the adjusted profile likelihood of~\citet{mccullagh1990simple}, we estimate $\tau^2$ by solving the centered profile score equation, that is, $U^{\mathrm{PL}}_{\tau^2}(\tau^2) - \boldsymbol{E}\{U^{\mathrm{PL}}_{\tau^2}(\tau^2)\}=0$. Since the exact expectation of the profile score is difficult to derive, we replace it with its approximation, denoted as $m(\xi)$. The details of $m(\xi)$ are given in Section~\ref{sec.3.3} and Web Appendix A. Thus, we solve
\begin{equation}
    U^{\mathrm{PL}}_{\tau^2}(\tau^2)-m\{\hat{\theta}(\tau^2),\tau^2,\hat{\beta}(\tau^2)\}=0.
    \label{eq.15}
\end{equation}
We denote the solution by $\hat{\tau}^2_{\REMLadj}$; the estimates of $\theta$ and $\beta$ are $\hat{\theta}_{\REMLadj}=\hat{\theta}(\hat{\tau}^2_{\REMLadj})$ and $\hat{\beta}_{\REMLadj}=\hat{\beta}(\hat{\tau}^2_{\REMLadj})$. Since $U_\theta=U_\beta=0$ at $\{\hat{\theta}(\tau^2),\tau^2,\hat{\beta}(\tau^2)\}$, solving~\eqref{eq.15} is equivalent to solving the joint system
\begin{equation}
    U_{\theta}(\xi)=0,\qquad U_{\tau^2}(\xi)-m(\xi)=0,\qquad U_{\beta}(\xi)=0.
    \label{eq.16}
\end{equation}
In~\eqref{eq.16}, the first and third equations, $U_\theta=0$ and $U_\beta=0$, are the score equations of $\ell_{\mathrm{cond}}$ in~\eqref{eq.12}, which are the same as those solved by the MLE; only the second equation, for $\tau^2$, differs from that of the MLE by the adjustment term $m(\xi)$. In the absence of PB, \eqref{eq.15} reduces exactly to the REML equation~\eqref{eq.6}, that is, $\hat{\tau}^2_{\REMLadj}=\hat{\tau}^2_{\REMLunadj}$; the proof is given in Web Appendix B. Thus, this estimator extends the REML to the sensitivity analysis for PB. The estimate of $\alpha$ is given by $\hat{\alpha}$ evaluated at the estimates of $(\theta, \tau^2, \beta)$, that is, $\hat{\alpha}_{\REMLadj}=\hat{\alpha}(\hat{\theta}_{\REMLadj},\hat{\tau}^2_{\REMLadj}, \hat{\beta}_{\REMLadj})$.

\subsection{Bias adjustment of the profile score}\label{sec.3.3}

We now construct $m(\xi)$. We use the cumulant notation of~\citet{mccullagh1987tensor}. For $j, k, l\in\{\theta, \tau^2, \beta\}$, $U_j$ is the score function in~\eqref{eq.12}, and additional subscripts of $U$ denote derivatives with respect to the corresponding parameters, for example, $U_{\tau^2\theta}=\partial U_{\tau^2}/\partial\theta$. Let $\kappa_{j,k}=\boldsymbol{E}(U_jU_k)$, $\kappa_{jk,l}=\boldsymbol{E}(U_{jk}U_l)$, and $\kappa_{jkl}=\boldsymbol{E}(U_{jkl})$, where the expectations are taken under the conditional density of published studies, and let $\kappa^{r,s}$ denote the elements of the inverse of the $2\times 2$ block $(\kappa_{r,s})$ for $\lambda$, with the indices $r, s, t, u$ ranging over $\{\theta, \beta\}$; $\lambda_r$ and $\hat{\lambda}_r$ denote the elements of $\lambda$ and $\hat{\lambda}$ corresponding to $r$, that is, $\lambda_\theta=\theta$, $\lambda_\beta=\beta$, $\hat{\lambda}_\theta=\hat{\theta}(\tau^2)$, and $\hat{\lambda}_\beta=\hat{\beta}(\tau^2)$. Expanding the profile score around the true $\lambda$ and taking the expectation of each term, we obtain
\begin{equation}
    \boldsymbol{E}\{U^{\mathrm{PL}}_{\tau^2}(\tau^2)\} = m(\xi) + o(1),\quad m(\xi)=-\sum_{r} \kappa_{\tau^2,r}\,\delta_r + \sum_{r,s} \kappa^{r,s}\kappa_{\tau^2 r,s} + \frac{1}{2}\sum_{r,s} \kappa^{r,s}\kappa_{\tau^2 rs},
    \label{eq.17}
\end{equation}
where $\delta_r=\sum_{s,t,u} \kappa^{r,s}\kappa^{t,u}(\kappa_{stu}/2+\kappa_{st,u})$ is the first-order bias of $\hat{\lambda}_r$~\citep{cox1968general}. Since all the cumulants are expectations under the conditional density of a published study, $m(\xi)$ is a known function of $\xi$, and in~\eqref{eq.15} it is evaluated at $\{\hat{\theta}(\tau^2),\tau^2,\hat{\beta}(\tau^2)\}$. The derivation of~\eqref{eq.17} is given in Web Appendix A.

\subsection{A Jeffreys-type penalty for small numbers of studies}\label{sec.3.4}

When the number of published studies is small, $\ell_{\mathrm{cond}}$ can be monotone in $\beta$, and the estimate of $\beta$ can diverge. In the MLE, this is handled by bounding $\beta$ within a pre-specified range, and the estimate of $\beta$ often lies on the bound in small meta-analyses. To stabilize the estimation of $\beta$, we add the Jeffreys-type penalty $\mathrm{pen}(\xi)=-\frac{1}{2}\log\{[K(\xi)^{-1}]_{\beta\beta}\}$ to the estimating equations of $\theta$ and $\beta$~\citep{firth1993bias,kosmidis2021jeffreys}, where $K(\xi)=(\kappa_{j,k})$ is the $3\times 3$ expected information matrix. The penalty tends to $-\infty$ as $|\beta|\to\infty$ and thus keeps the estimate of $\beta$ finite. That is, for a given $\tau^2$, $\hat{\theta}(\tau^2)$ and $\hat{\beta}(\tau^2)$ now solve
\begin{equation}
    U_{\theta}(\xi)+\frac{\partial\,\mathrm{pen}(\xi)}{\partial \theta}=0,\qquad U_{\beta}(\xi)+\frac{\partial\,\mathrm{pen}(\xi)}{\partial \beta}=0.
    \label{eq.18}
\end{equation}
Since the penalty changes the bias of $\hat{\lambda}$, the first part of $m(\xi)$ is modified accordingly. Our final estimating equation for $\tau^2$ is
\begin{equation}
    U^{\mathrm{PL}}_{\tau^2}(\tau^2)-\tilde{m}\{\hat{\theta}(\tau^2),\tau^2,\hat{\beta}(\tau^2)\}=0,\qquad
    \tilde{m}(\xi)=m(\xi)-\sum_{r,s} \kappa_{\tau^2,r}\,\kappa^{r,s}\,\frac{\partial\,\mathrm{pen}(\xi)}{\partial \lambda_s}.
    \label{eq.19}
\end{equation}
Equivalently, $\xi$ solves the joint system
\begin{equation}
    U_{\theta}(\xi)+\frac{\partial\,\mathrm{pen}(\xi)}{\partial \theta}=0,\qquad U_{\tau^2}(\xi)-\tilde{m}(\xi)=0,\qquad U_{\beta}(\xi)+\frac{\partial\,\mathrm{pen}(\xi)}{\partial \beta}=0.
    \label{eq.20}
\end{equation}
We denote the solution of~\eqref{eq.20} by $\hat{\xi}=(\hat{\theta}_{\mathrm{Proposed}},\hat{\tau}^2_{\mathrm{Proposed}},\hat{\beta}_{\mathrm{Proposed}})$.
Without PB, there is no $\beta$ and no penalty, and~\eqref{eq.19} is still the REML equation, that is, $\hat{\tau}^2_{\mathrm{Proposed}}=\hat{\tau}^2_{\REMLunadj}$. A brief illustration of how the penalty works is given in Web Appendix C.

\subsection{Computation and inference}\label{sec.3.5}

All expectations in $m(\xi)$, $\tilde{m}(\xi)$, and $\mathrm{pen}(\xi)$ are one-dimensional integrals with respect to the conditional density of a published study given $\sigma_i$, which we compute by 40-point Gauss-Hermite quadrature. The third-order cumulants are obtained from the Bartlett identities~\citep{bartlett1953approximate,mccullagh1987tensor}, so that no third derivatives are required. We solve the joint system~\eqref{eq.20} by a damped Newton method starting from the penalized MLE. The details of the computation are given in Web Appendix D.

The estimator solving~\eqref{eq.20} is a $Z$-estimator. We estimate the variance of $\hat{\xi}$ by the sandwich estimator $\hat{A}^{-1}\hat{B}\hat{A}^{-T}$~\citep{kauermann2001note}, where $\hat{A}$ is the derivative of the estimating functions in~\eqref{eq.20} at $\hat{\xi}$ and $\hat{B}$ is the sum of the outer products of the study-level estimating functions (Web Appendix D). Based on the asymptotic normality of $\hat{\xi}$, we construct the Wald-type 95\% CIs for $\theta$ and $\tau^2$ as $\hat{\theta}\pm 1.96\,\widehat{\mathrm{SE}}(\hat{\theta})$ and $[\max\{0,\hat{\tau}^2-1.96\,\widehat{\mathrm{SE}}(\hat{\tau}^2)\},\,\hat{\tau}^2+1.96\,\widehat{\mathrm{SE}}(\hat{\tau}^2)]$, respectively, where $\hat{\tau}^2$ is $\hat{\tau}^2_{\mathrm{Proposed}}$, $\hat{\theta}$ is $\hat{\theta}_{\mathrm{Proposed}}$, and the standard errors are the square roots of the corresponding diagonal elements of the sandwich estimator. For the MLE, we use the same sandwich estimator with the score functions of $\ell_{\mathrm{cond}}$ and $(\hat{\theta},\hat{\tau}^2)=(\hat{\theta}_{\MLEadj},\hat{\tau}^2_{\MLEadj})$. When the MLE of $\beta$ lies on its bound, $\beta$ is treated as fixed at the bound in the sandwich estimator, so that the resulting CIs do not reflect the uncertainty in $\beta$.

To facilitate interpretation of the estimated between-study variance, we also calculate the global $I^2$ statistic with its 95\% CI~\citep{higgins2002quantifying}. Let $W_1=\sum_{i=1}^{N}1/\sigma_i^2$ and $W_2=\sum_{i=1}^{N}1/\sigma_i^4$. We define the typical within-study variance as $s_{\mathrm{typ}}^2=(N-1)W_1/(W_1^2-W_2)$. For any estimate $\hat{\tau}^2$, we calculate $\widehat{I}^2=\hat{\tau}^2/(\hat{\tau}^2+s_{\mathrm{typ}}^2)$ and report it as a percentage; because this transformation is monotone in $\hat{\tau}^2$, its 95\% CI is obtained by applying the same transformation to the confidence limits for $\tau^2$.

\section{Simulation studies}\label{sec.4}

We carried out simulation studies to evaluate the performance of the proposed heterogeneity estimator and to assess how it affects the estimation of the common effect size $\theta$.

We set the common effect size $\theta=2$ and the between-study heterogeneity $\tau=0.25, 0.5, 0.75$, and $1$ to simulate meta-analyses with four different heterogeneity levels. The median within-study standard error was set to two levels, $\bar{\sigma}=0.3$ and $0.5$. This yielded a total of 8 ($4\times 2$) settings. The number of published studies $S$ was set to 15, 25, 50, and 200. A meta-analysis with 200 published studies may seem implausible, but we included this setting to examine the large-sample properties. For each study, the true effect size was sampled as $\theta_i \sim N(\theta, \tau^2)$, the within-study standard error $\sigma_i$ was sampled from the log-normal distribution $LN(\log \bar{\sigma}, 0.1^2)$, and the observed effect size was sampled as $\hat{\theta}_i \sim N(\theta_i, \sigma_i^2)$. We decided whether a study was published with a Bernoulli distribution, that is, $\mathrm{Bern}\{\Phi(\alpha+\beta\hat{\theta}_i/\sigma_i)\}$, with $\beta=1$. The parameter $\alpha$ was determined for each setting so that the marginal selection probability was 0.7, meaning approximately 70\% of studies were published; it was computed by Monte Carlo integration with $2\times 10^6$ draws. Studies were generated one at a time until $S$ studies were published. We referred to all generated studies as the population dataset and to the published studies as the published dataset. Under each setting, we generated 1000 independent meta-analyses.

We applied the proposed estimator (Proposed) and the MLE of~\citet{copas2013likelihood} (MLE$_{\mathrm{adj}}$) to the published datasets with the true $p=0.7$; for MLE$_{\mathrm{adj}}$, $\beta$ was bounded by $\pm 10/\{\max_i(\hat{\theta}_i)-\min_i(\hat{\theta}_i)\}$ as in the original implementation. We also applied the ordinary ML and REML to the published datasets without accounting for PB (MLE$_{\mathrm{pub}}$ and REML$_{\mathrm{pub}}$) to quantify the magnitude of PB, and to the population datasets (MLE$_{\mathrm{pop}}$ and REML$_{\mathrm{pop}}$), which served as the ideal estimators in the absence of PB. The ML and REML fits were obtained with the \pkg{R} package \pkg{metafor}~\citep{viechtbauer2010conducting}, with the Wald-type CI for $\theta$ and the Q-profile CI for $\tau^2$~\citep{viechtbauer2007confidence}. The performance was summarized over the converged replicates; the convergence rates were 96\%--100\% for both MLE$_{\mathrm{adj}}$ and Proposed.

\begin{figure}[htb]
    \centering
    \includegraphics[width=\linewidth,height=0.8\textheight,keepaspectratio]{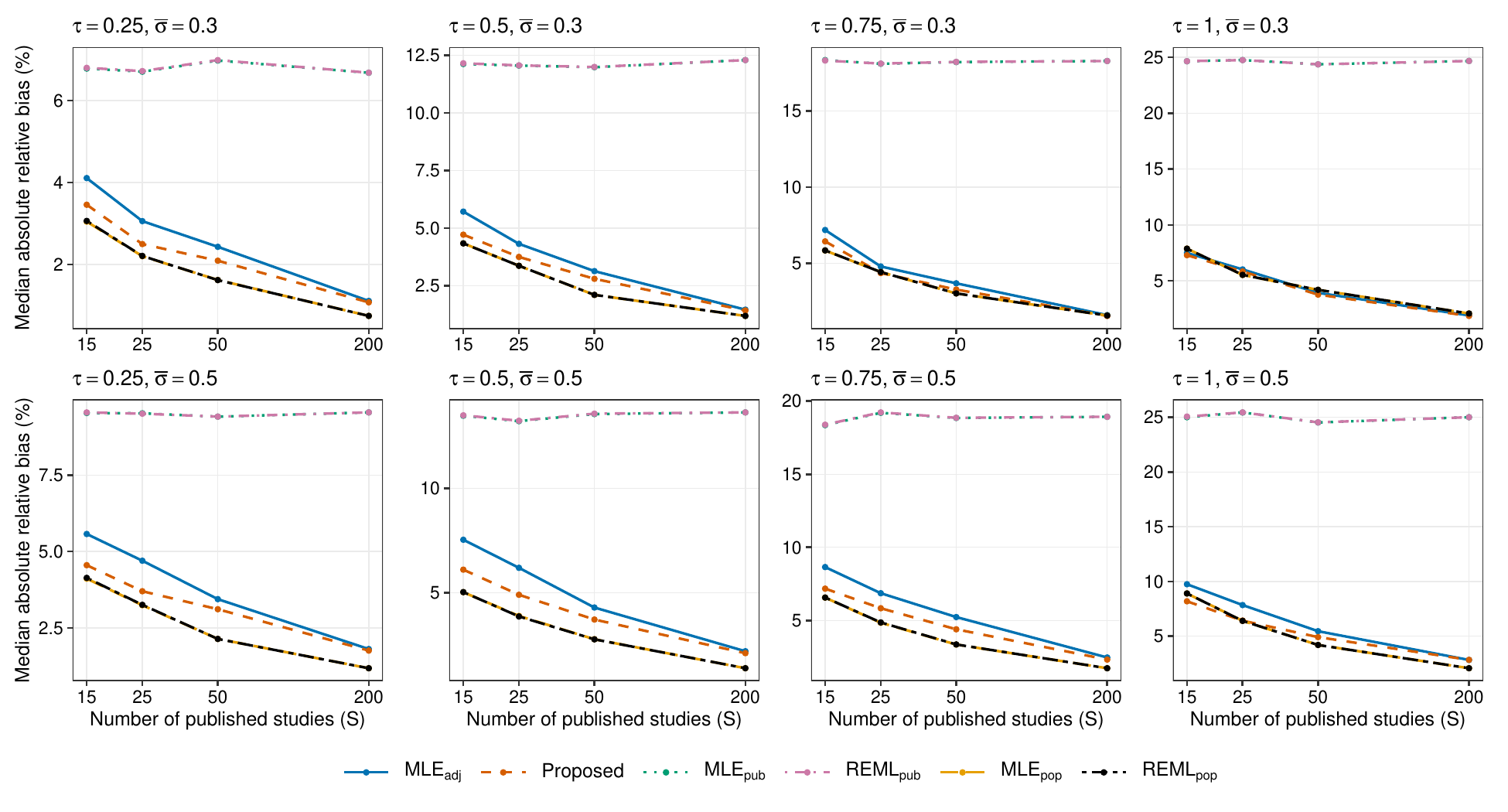}
    \caption{Median absolute relative bias (MedARB, \%) of the estimated $\theta$ against the number of published studies $S$. Top row: $\bar{\sigma}=0.3$; bottom row: $\bar{\sigma}=0.5$; columns: $\tau=0.25, 0.5, 0.75, 1$. MLE$_{\mathrm{adj}}$, the MLE of~\citet{copas2013likelihood}; Proposed, the proposed estimator; MLE$_{\mathrm{pub}}$ and REML$_{\mathrm{pub}}$, ML and REML on the published studies without adjustment for PB; MLE$_{\mathrm{pop}}$ and REML$_{\mathrm{pop}}$, ML and REML on all generated studies.}
    \label{fig:fig1}
\end{figure}

\begin{figure}[htb]
    \centering
    \includegraphics[width=\linewidth,height=0.8\textheight,keepaspectratio]{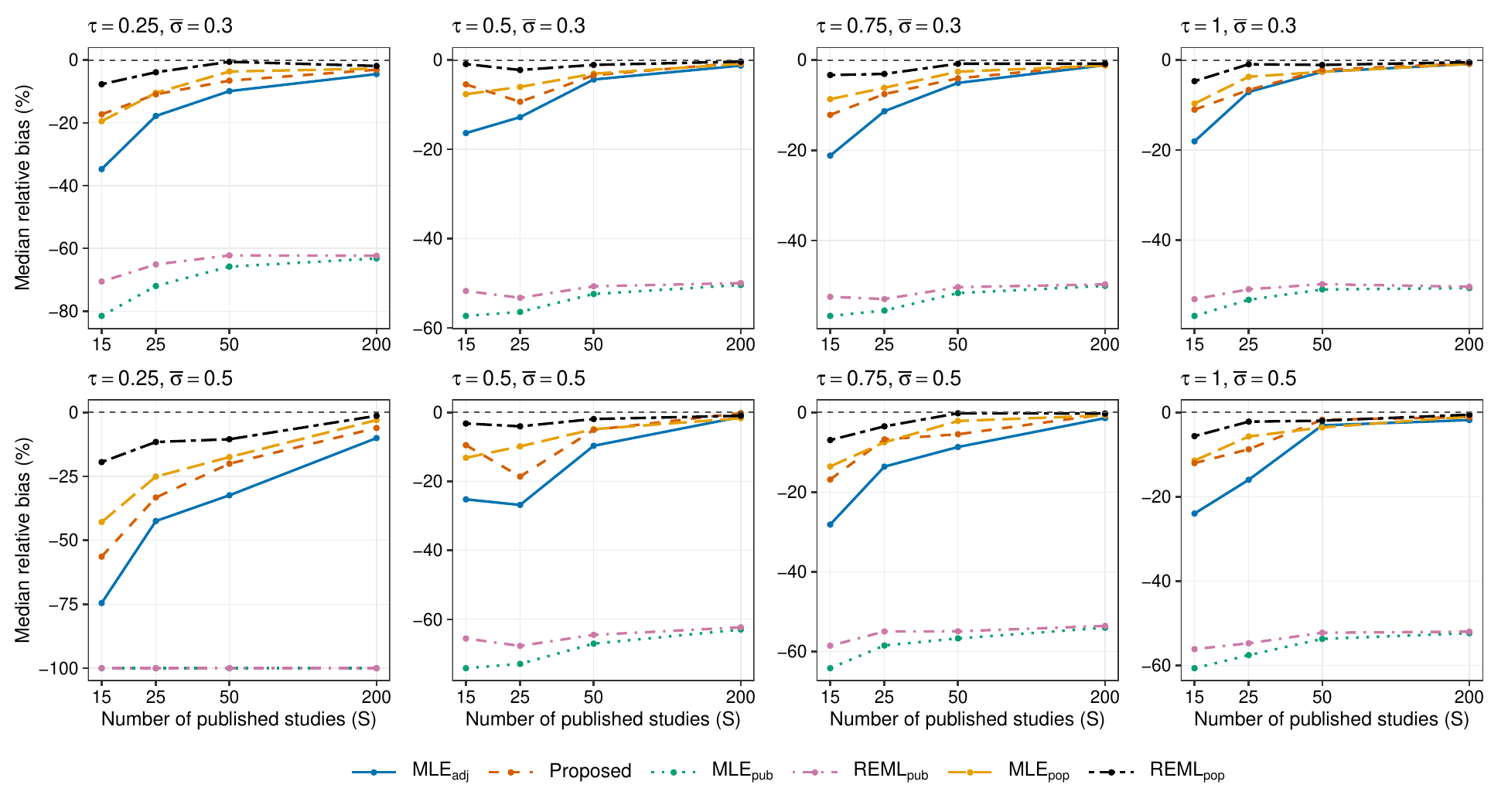}
    \caption{Median relative bias (MedRB, \%) of the estimated $\tau^2$ against the number of published studies $S$. The layout of the panels and the methods are the same as in Figure~\ref{fig:fig1}.}
    \label{fig:fig2}
\end{figure}

Since the estimates of $\tau^2$ have a skewed distribution with a point mass at 0 when $S$ is small, we summarize the bias for $\tau^2$ by the median relative bias (MedRB), $\operatorname{Median}_g\{(\hat{\tau}^2_g-\tau^2)/\tau^2\}$, where $\hat{\tau}^2_g$ is the estimate from the $g$-th meta-analysis~\citep{morris2019using}. For $\theta$, we summarize the estimation error by the median absolute relative bias (MedARB), $\operatorname{Median}_g\{|\hat{\theta}_g-\theta|/\theta\}$, which reflects both bias and variability. We show the MedARB for $\theta$ and the MedRB for $\tau^2$ in Figures~\ref{fig:fig1} and~\ref{fig:fig2}, respectively. The mean squared error (MSE), the empirical coverage probability and mean length of the 95\% CIs, and the convergence rates for $\theta$ and $\tau^2$ are provided in Web Appendix E.

We observe in Figure~\ref{fig:fig1} that the MedARB for $\theta$ of Proposed was smaller than that of MLE$_{\mathrm{adj}}$ in 31 of the 32 combinations of settings and $S$, and equal in the remaining one. MLE$_{\mathrm{pub}}$ and REML$_{\mathrm{pub}}$ had errors of 6.7\%--25.4\% that did not decrease with $S$, which reflects the bias due to PB. For $\tau^2$, both MLE$_{\mathrm{adj}}$ and Proposed had a negative MedRB when $S$ was small (Figure~\ref{fig:fig2}), which partly reflects the right-skewed distribution of $\hat{\tau}^2$ with a point mass at 0; even REML$_{\mathrm{pop}}$ also had a negative MedRB at $S=15$ in all settings. Proposed had a smaller absolute MedRB than MLE$_{\mathrm{adj}}$ in all 32 combinations for $\tau^2$; at $S=15$, the MedRB ranged from $-74.5\%$ to $-16.4\%$ for MLE$_{\mathrm{adj}}$ and from $-56.4\%$ to $-5.4\%$ for Proposed. The difference between the two estimators decreased as $S$ increased, which is consistent with their asymptotic equivalence. In contrast, MLE$_{\mathrm{pub}}$ and REML$_{\mathrm{pub}}$ had a non-vanishing MedRB between $-100\%$ and $-49.7\%$.

The improvement in $\tau^2$ carried over to the inference on $\theta$. At $S\le 50$, the coverage probabilities of the 95\% CIs for $\theta$ were 84\%--92\% for Proposed, compared with 69\%--86\% for MLE$_{\mathrm{adj}}$; for $\tau^2$, they were 78\%--98\% and 70\%--97\%, respectively. Both approached the nominal level at $S=200$. The estimate of $\beta$ by MLE$_{\mathrm{adj}}$ lay on its bound in 25.5\%--52.6\% of the replicates at $S=15$, whereas that by Proposed lay on the bound in at most 2.9\% of the replicates under all settings (Web Appendix C). In these replicates, the CIs of MLE$_{\mathrm{adj}}$ were computed with $\beta$ fixed at the bound, as described in Section~\ref{sec.3.5}. Proposed had a slightly larger MSE for $\tau^2$ than MLE$_{\mathrm{adj}}$ at $S=15$ and 25 (ratios 1.00--1.26), and a comparable MSE at $S\ge 50$; a similar relation is seen between REML$_{\mathrm{pop}}$ and MLE$_{\mathrm{pop}}$ (ratios 1.00--1.23), which reflects the usual trade-off between the bias and the variance of the REML.

\section{Applications}\label{sec.5}

We illustrated the proposed method with a meta-analysis conducted by~\citet{chen2013high} to evaluate the effectiveness of the high maintenance-dose clopidogrel in reducing major adverse cardiovascular\slash cerebrovascular events compared with the standard dose. This is a small-scale meta-analysis with only 12 published studies, a typical setting in which the ML underestimates $\tau^2$. The data are given in Table~\ref{tab:data_clopi}. The effect size of interest is the log odds ratio (lnOR). The pooled OR was reported as 0.60 (95\% CI: [0.43, 0.83]) in the original paper, suggesting strong effectiveness of the high maintenance-dose clopidogrel. However, the data may be subject to PB, with the $p$-value of Macaskill's test equal to 0.02. We therefore adjusted for PB to assess how PB might influence the inference. We applied the MLE of~\citet{copas2013likelihood} and our proposed method to estimate the common lnOR, the between-study variance, and the global $I^2$ under $p=1, 0.9, \cdots, 0.2$, avoiding $p=0.1$ as advised by~\citet{copas2013likelihood} since bias correction under small $p$ implies excessive extrapolation. Since a smaller lnOR indicates a beneficial effect, $\beta$ was restricted to be non-positive so that studies with a more beneficial effect were more likely published. Note that $p=1$ corresponds to the case of no PB, and the estimates under $p=1$ are obtained from the conventional ML or REML. The marginal selection probability has a one-to-one mapping to the number of unpublished studies $M=N/p-N$. Thus, the sensitivity analysis results can be interpreted in terms of the number of unpublished studies. We show the results in Table~\ref{tab:clopi}.

\begin{table}[htb]
    \centering
    \begin{threeparttable}
    \caption{Data of the clopidogrel meta-analysis.}
    \label{tab:data_clopi}
    \begin{tabular}{cccccc}
\hline \multirow{2}{*}{Study} & \multicolumn{2}{c}{High dose} & \multicolumn{2}{c}{Standard dose} & \multirow{2}{*}{Total subjects} \\
\cline{2-5} & Events & Subjects & Events & Subjects & \\
\hline
1 & 1 & 36 & 8 & 38 & 74 \\
2 & 0 & 24 & 1 & 24 & 48 \\
3 & 2 & 47 & 8 & 47 & 94 \\
4 & 25 & 1109 & 25 & 1105 & 2214 \\
5 & 1 & 21 & 2 & 23 & 44 \\
6 & 4 & 403 & 9 & 410 & 813 \\
7 & 6 & 46 & 10 & 55 & 101 \\
8 & 4 & 205 & 2 & 195 & 400 \\
9 & 1 & 30 & 2 & 30 & 60 \\
10 & 0 & 58 & 1 & 62 & 120 \\
11 & 1 & 31 & 1 & 29 & 60 \\
12 & 14 & 150 & 30 & 156 & 306 \\
\hline
\end{tabular}
    \begin{tablenotes}[flushleft]\footnotesize
    \item Note: Events, number of patients with major adverse cardiovascular/cerebrovascular events; Subjects, number of patients in the arm; Total subjects, sum of the two arms.
    \end{tablenotes}
    \end{threeparttable}
\end{table}

\begin{table}[htb]
    \centering
    \begin{threeparttable}
    \caption{Sensitivity analysis of the clopidogrel meta-analysis ($N=12$) under the marginal selection probability $p$.}
    \label{tab:clopi}
    \small
    \begin{tabular}{ccccc}
\hline \multicolumn{5}{c}{MLE based} \\
\hline $p$ & \# of unpublished & $\theta$ & $\tau^2$ & $I^2(\%)$ \\
\hline 1.0 & 0 & -0.527[-0.930,-0.123] & 0.058[0.000,0.754] & 11.7[0.0,63.3] \\
0.9 & 1 & -0.448[-0.813,-0.082] & 0.079[0.000,0.189] & 15.3[0.0,30.2] \\
0.8 & 3 & -0.369[-0.775,0.036] & 0.092[0.000,0.220] & 17.4[0.0,33.5] \\
0.7 & 5 & -0.262[-0.680,0.157] & 0.125[0.000,0.302] & 22.2[0.0,40.8] \\
0.6 & 8 & -0.137[-0.632,0.358] & 0.172[0.000,0.440] & 28.2[0.0,50.2] \\
0.5 & 12 & 0.010[-0.705,0.725] & 0.232[0.000,0.718] & 34.7[0.0,62.1] \\
0.4 & 18 & 0.188[-0.924,1.300] & 0.307[0.000,1.166] & 41.2[0.0,72.7] \\
0.3 & 28 & 0.406[-1.203,2.016] & 0.393[0.000,1.691] & 47.3[0.0,79.5] \\
0.2 & 48 & 0.689[-1.347,2.724] & 0.490[0.000,2.117] & 52.8[0.0,82.9] \\
\hline
\end{tabular}

    \vspace{6pt}

    \begin{tabular}{ccccc}
\hline \multicolumn{5}{c}{Proposed} \\
\hline $p$ & \# of unpublished & $\theta$ & $\tau^2$ & $I^2(\%)$ \\
\hline 1.0 & 0 & -0.547[-0.980,-0.114] & 0.095[0.000,0.754] & 17.9[0.0,63.3] \\
0.9 & 1 & -0.508[-0.904,-0.113] & 0.169[0.000,0.380] & 27.9[0.0,46.5] \\
0.8 & 3 & -0.406[-0.820,0.007] & 0.193[0.000,0.418] & 30.6[0.0,48.9] \\
0.7 & 5 & -0.285[-0.734,0.164] & 0.225[0.000,0.477] & 34.0[0.0,52.2] \\
0.6 & 8 & -0.147[-0.661,0.367] & 0.268[0.000,0.605] & 38.0[0.0,58.0] \\
0.5 & 12 & 0.009[-0.605,0.623] & 0.321[0.000,0.791] & 42.4[0.0,64.4] \\
0.4 & 18 & 0.194[-0.528,0.917] & 0.394[0.000,1.079] & 47.4[0.0,71.2] \\
0.3 & 28 & 0.427[-0.593,1.446] & 0.478[0.000,1.428] & 52.2[0.0,76.6] \\
0.2 & 48 & 0.705[-0.470,1.879] & 0.562[0.000,1.636] & 56.2[0.0,78.9] \\
\hline
\end{tabular}
    \begin{tablenotes}[flushleft]\footnotesize
    \item Note: \# of unpublished, estimated number of unpublished studies; $\theta$, common lnOR; $\tau^2$, between-study variance; $I^2$, global $I^2$. Each cell shows the estimate with the 95\% CI in brackets. At $p=1$, the rows are the ML (MLE based) and REML (Proposed) estimates with the Wald-type CI for $\theta$ and the Q-profile CI for $\tau^2$; at $p<1$, the CIs are sandwich-based Wald CIs.
    \end{tablenotes}
    \end{threeparttable}
\end{table}

The estimated lnOR increased as the number of unpublished studies grew, and both methods indicated that the significant effect could be overturned with as few as three unpublished studies ($p=0.8$). Thus, the significant effect of the high maintenance-dose clopidogrel should be interpreted with caution. The estimates of $\tau^2$ by our method were larger than those by the MLE under every $p$. In the absence of selective publication ($p = 1$), the ML estimate of $\tau^2$ was 0.058 while the REML estimate was 0.095, corresponding to $I^2$ values of 11.7\% and 17.9\%, respectively. Thus, the MLE-based sensitivity analysis started from an estimate 39\% smaller than the REML estimate commonly reported in the primary analysis, whereas our method started from the REML estimate. In the presence of selective publication, the estimate of $\tau^2$ increased with the number of unpublished studies up to 0.490 ($I^2=52.8\%$) with the MLE and up to 0.562 ($I^2=56.2\%$) with our method. At $p=0.9$ with only one unpublished study, the estimate of $\tau^2$ by our method (0.169) was about twice that by the MLE (0.079), and the corresponding $I^2$ was 27.9\% and 15.3\%, respectively. This is consistent with the simulation results, in which the MLE underestimated $\tau^2$ in small meta-analyses. The estimates of the common lnOR were close between the two methods. The CIs for $\theta$ from our method were narrower than those from the MLE when $p\le 0.5$.

We also applied both methods to a larger meta-analysis of nicotine gum for smoking cessation with 56 published studies~\citep{stead2012nicotine}. The estimates by the two methods were close under every $p$, which is consistent with their asymptotic equivalence when the number of studies is large. With our method, the significant effect vanished at $p=0.6$. With the MLE, the 95\% CI for the common lnOR included 0 from $p=0.7$. However, the MLE of $\beta$ lay on its bound at $p=0.9$ and $0.8$, so that the CIs at these $p$ did not reflect the uncertainty in $\beta$, and at $p=0.7$ the likelihood was nearly flat in $\beta$, which led to wide CIs for all parameters. In contrast, our estimate of $\beta$ was an interior point under all $p$, and the CIs changed smoothly with $p$. The details are given in Web Appendix F.

\section{Discussion}\label{sec.6}

In this work, we proposed a REML-type estimator of the between-study variance for the selection-function-based sensitivity analysis for PB. The existing sensitivity analysis methods estimate $\tau^2$ by the ML conditional on published studies, and the REML, the recommended estimator without PB, cannot be directly applied since error contrasts free of the common effect size are not available under selective publication. However, the REML can also be constructed by centering the profile score for $\tau^2$ by its mean. Following this observation, the proposed method extends the REML to the sensitivity analysis for PB by centering the profile score for $\tau^2$ by its approximate mean. To the best of our knowledge, it serves as the first REML-type method in the sensitivity analysis for PB, and it reduces exactly to the REML without PB. The improvement was not limited to $\tau^2$. Since $\tau^2$ determines the weights of the studies, the proposed estimator also gave smaller estimation errors and better coverage probabilities for $\theta$ than the MLE in our simulation studies, especially when the number of published studies was small. Since most meta-analyses include a small number of studies, the proposed method would be a more accurate alternative to the ML-based sensitivity analysis in practice. In addition, the Jeffreys-type penalty kept the estimate of the selection parameter finite, whereas the MLE of the selection parameter frequently lay on its bound in small meta-analyses, in which case its CIs do not reflect the uncertainty in the selection parameter.

Our method belongs to the class of sensitivity analysis methods that rely on sensitivity parameters. Other approaches use weighted estimators or estimate the selection function by incorporating information on unpublished studies from clinical registries~\citep{huang2021using,huang2023adjusting}. \citet{huang2023adjusting} proposed extended DerSimonian-Laird estimators by incorporating estimated selection functions; their method requires auxiliary information on unpublished studies from clinical trial registries, while the adjustment in~\eqref{eq.17} requires only the conditional density of a published study and the moments of its score functions.

Though our method showed good performance in both real-data analysis and simulation studies, there remain some limitations. First, our estimator follows the sensitivity analysis method of~\citet{copas2013likelihood} and assumes a two-parameter probit selection function based on the $t$-statistic. Many other parametric or nonparametric selection functions have been proposed to address PB~\citep{copas2004bound,zhou2024nonparametric,huang2023adjusting}. Although the adjustment can be applied to other selection functions in principle, the computation of the expectations may be considerably more complex, for example, for selection functions without a closed-form marginal selection probability. Second, we estimate the parameters under the conditional likelihood with a given marginal selection probability $p$ as the sensitivity parameter. \citet{copas2013likelihood} proposed this constraint approach to reduce the number of parameters, while the identifiability issue remains to be examined from a theoretical perspective. Third, our adjustment centers the profile score for $\tau^2$, which removes the bias of $\hat{\tau}^2$ caused by estimating $\theta$ and $\beta$, as the REML does in the NN model. However, when $S$ is small, the sampling distribution of $\hat{\tau}^2$ is right-skewed with a point mass at 0. In our simulation studies, the MedRB of our estimator at $S=15$ was negative and non-negligible in all settings; REML$_{\mathrm{pop}}$ showed the same pattern without PB. Our estimator also had a slightly larger MSE for $\tau^2$ than the MLE at $S\le 25$, and the coverage probabilities of the Wald-type CIs were below the nominal level for small $S$. It is appealing to develop CIs with better small-sample coverage, such as profile-type intervals, in future work. In addition, our estimator is developed under the NN model, which may be unrealistic in rare-event meta-analyses, so our estimator is expected to inherit this drawback in such settings. A number of methods have been proposed for rare-event meta-analysis, including generalized linear mixed models and the beta-binomial model~\citep{stijnen2010random,kuss2015statistical}, and some have been extended to the case of PB~\citep{hu2024sensitivity}. It is appealing to extend our estimator to address rare events in future work.

\section*{Acknowledgements}
This research was partly supported by Grant-in-Aid for Challenging Exploratory Research (25K03086, 26K21181) from the Ministry of Education, Science, Sports and Technology of Japan. We thank Claude Opus 5 (Anthropic) for assisting in developing a faster and more robust algorithm for the numerical approximation of the expectations in the proposed adjustment.

\section*{Conflict of Interest}
The authors have declared no conflict of interest.

\section*{Data Availability Statement}
The data that support the findings of this study are provided in Table~\ref{tab:data_clopi} and Web Appendix F of the Supporting Information. The \pkg{R} code for the proposed method, the simulation studies, and the data analyses is provided as Supporting Information.

\section*{Ethics Statement}
This study used only summary data from published studies, and ethical approval was not required.

\section*{Declaration of Generative Artificial Intelligence Use}
In this work, we used the generative artificial intelligence model Claude Opus 5 (Anthropic) in two ways. First, it was used as a writing aid to suggest alternative word choices, refine sentence structure, explore alternative phrasings for ideas within paragraphs, and solve formatting issues. Second, it assisted in developing the algorithm for the numerical approximation of the expectations in the proposed adjustment and in writing and checking the \pkg{R} code. All conceptual contributions, study design, and final writing decisions were made by the authors. The authors verified the algorithm, the code, and all results, and take full responsibility for the accuracy and originality of the content.

%======================================================================
% Supporting Information
%======================================================================
\newpage
\setcounter{page}{1}
\setcounter{section}{0}
\setcounter{equation}{0}
\setcounter{table}{0}
\setcounter{figure}{0}
\renewcommand{\thetable}{S\arabic{table}}
\renewcommand{\thefigure}{S\arabic{figure}}
\renewcommand{\theequation}{S\arabic{equation}}
\renewcommand{\thesection}{Web Appendix \Alph{section}}
\renewcommand{\thesubsection}{\Alph{section}.\arabic{subsection}}
\renewcommand{\theHtable}{S\arabic{table}}
\renewcommand{\theHfigure}{S\arabic{figure}}
\renewcommand{\theHequation}{S\arabic{equation}}
\renewcommand{\theHsection}{S\Alph{section}}
\renewcommand{\theHsubsection}{S\Alph{section}.\arabic{subsection}}

\begin{center}
{\LARGE Supporting Information for ``Improving the selection-function-based sensitivity analysis for publication bias with restricted maximum likelihood''\par}
\vskip 1.5em
{\large Taojun Hu, Yi Zhou, Xiao-Hua Zhou, and Satoshi Hattori\par}
\end{center}
\vskip 1em

\noindent Equation, section, and table numbers without the prefix ``S'' refer to the main text.

%======================================================================
\section{Score functions and the approximate mean of the profile score}\label{ap:A}

\subsection{Score functions}

Throughout this appendix, $\hat{\alpha}=\hat{\alpha}(\theta,\tau^2,\beta)$ is the solution of the constraint~(\ref{eq.9}), $\gamma_{1i} = \hat{\alpha} + \beta\theta/\sigma_i$, $\gamma_{2i} = \sqrt{ 1 + \beta^2(1 + \tau^2/\sigma_i^2) }$, $z_i=\gamma_{1i}/\gamma_{2i}$, $\eta_i=\hat{\alpha} + \beta \hat{\theta}_i/\sigma_i$, and $\varphi(\cdot)$ is the probability density function of the standard normal distribution. The score functions of the conditional log-likelihood $\ell_{\mathrm{cond}}$ in~(\ref{eq.11}) with respect to $\xi=(\theta,\tau^2,\beta)$ are
\begin{equation}
\label{eq.S1}
\begin{aligned}
    U_\theta &= \sum_{i=1}^N \left[ \frac{\hat{\theta}_i-\theta}{\tau^2 + \sigma_i^2} - \frac{\varphi(z_i)}{\Phi(z_i)}\frac{\beta}{\sigma_i \gamma_{2i}} + \left\{ \frac{\varphi(\eta_i)}{\Phi(\eta_i)} - \frac{\varphi(z_i)}{\Phi(z_i)}\frac{1}{\gamma_{2i}} \right\}\frac{\partial \hat{\alpha}}{\partial \theta}  \right],\\
    U_{\tau^2} &= \frac{1}{2}\sum_{i=1}^N \left\{ \frac{(\hat{\theta}_i -\theta)^2}{(\tau^2+\sigma_i^2)^2} - \frac{1}{\tau^2 + \sigma_i^2} + \frac{\varphi(z_i)}{\Phi(z_i)}\frac{\gamma_{1i} \beta^2}{\sigma_i^2\gamma_{2i}^3} \right\} + \sum_{i=1}^N \left\{ \frac{\varphi(\eta_i)}{\Phi(\eta_i)} - \frac{\varphi(z_i)}{\Phi(z_i)}\frac{1}{\gamma_{2i}} \right\}\frac{\partial \hat{\alpha}}{\partial \tau^2}, \\
    U_\beta &= \sum_{i=1}^N \left[ \frac{\varphi(\eta_i)}{\Phi(\eta_i)} \frac{\hat{\theta}_i}{\sigma_i} - \frac{\varphi(z_i)}{\Phi(z_i)}\frac{ \theta/\sigma_i - \hat{\alpha}\beta (1 + \tau^2/\sigma_i^2) }{\gamma_{2i}^3} + \left\{ \frac{\varphi(\eta_i)}{\Phi(\eta_i)} - \frac{\varphi(z_i)}{\Phi(z_i)}\frac{1}{\gamma_{2i}} \right\}\frac{\partial \hat{\alpha}}{\partial \beta} \right].
\end{aligned}
\end{equation}
Since $\ell_{\mathrm{cond}}$ depends on $\xi$ also through $\hat{\alpha}(\xi)$, each score includes the derivatives of $\hat{\alpha}$. Let $v_i=\varphi(z_i)/\Phi^2(z_i)$. Differentiating the constraint~(\ref{eq.9}) with respect to $(\theta,\tau^2,\beta)$ gives
\begin{equation*}
\begin{aligned}
    &\frac{\partial \hat{\alpha}}{\partial \theta} = -\frac{ \sum_{i=1}^N v_i\beta/(\sigma_i\gamma_{2i}) }{\sum_{i=1}^N v_i/\gamma_{2i}},\qquad
    \frac{\partial \hat{\alpha}}{\partial \tau^2} = \frac{1}{2}\frac{\sum_{i=1}^N v_i \gamma_{1i}\beta^2/(\sigma_i^2\gamma_{2i}^3)}{\sum_{i=1}^N v_i/\gamma_{2i}},\\
    &\frac{\partial \hat{\alpha}}{\partial \beta} = -\frac{\sum_{i=1}^N v_i\{\theta/\sigma_i -\hat{\alpha}\beta(1 + \tau^2/\sigma_i^2)\}/\gamma_{2i}^3 }{\sum_{i=1}^N v_i/\gamma_{2i}}.
\end{aligned}
\end{equation*}

Given $\sigma_i$, the conditional density of a published estimate is
\begin{equation*}
    g_i(y;\xi)=\frac{\Phi\{\hat{\alpha}(\xi)+\beta y/\sigma_i\}\,(\tau^2+\sigma_i^2)^{-1/2}\varphi\{(y-\theta)/(\tau^2+\sigma_i^2)^{1/2}\}}{\Phi(z_i)},
\end{equation*}
which integrates to one for every $\xi$. Since $\hat{\alpha}(\xi)$ depends only on $\xi$ and $(\sigma_1,\cdots,\sigma_N)$, the usual likelihood identities hold for $\ell_{\mathrm{cond},i}(\xi)=\log g_i(\hat{\theta}_i;\xi)$: the scores have mean zero and the Bartlett identities hold~\citep{S-bartlett1953approximate,S-mccullagh1987tensor}. Note that $\sum_{i=1}^N\ell_{\mathrm{cond},i}(\xi)$ equals $\ell_{\mathrm{cond}}(\xi)$ in~(\ref{eq.11}) up to terms free of the parameters.

\subsection{Derivation of the approximate mean}

We use the notation of Section~\ref{sec.3.3}: $\xi=(\xi_1,\xi_2,\xi_3)=(\theta,\tau^2,\beta)$, $\lambda=(\lambda_1,\lambda_2)=(\theta,\beta)$, $U_j$, $U_{jk}$, and $U_{jkl}$ for the first, second, and third derivatives of $\ell_{\mathrm{cond}}$ with respect to the parameters indicated by the indices $j,k,l\in\{\theta,\tau^2,\beta\}$, for example, $U_{\tau^2\theta}=\partial^2\ell_{\mathrm{cond}}/\partial\tau^2\partial\theta$, and
\begin{equation*}
\kappa_{j,k}=\boldsymbol{E}(U_jU_k),\quad \kappa_{jk,l}=\boldsymbol{E}(U_{jk}U_l),\quad \kappa_{j,k,l}=\boldsymbol{E}(U_jU_kU_l),\quad \kappa_{jkl}=\boldsymbol{E}(U_{jkl}),
\end{equation*}
with all expectations under $g_i$. Since the studies are independent and each study-level score has mean zero, each of these cumulants is the sum of the corresponding study-level cumulants, for example, $\kappa_{jk,l}=\sum_{i=1}^N\boldsymbol{E}\{(\partial^2\ell_{\mathrm{cond},i}/\partial\xi_j\partial\xi_k)(\partial\ell_{\mathrm{cond},i}/\partial\xi_l)\}$; we compute them in this form. The indices $r,s,t,u$ range over $\{\theta,\beta\}$, and $\lambda_r$ denotes the element of $\lambda$ corresponding to $r$, that is, $\lambda_\theta=\theta$ and $\lambda_\beta=\beta$, and $\kappa^{r,s}$ are the elements of the inverse of the block $(\kappa_{r,s})$ for $\lambda$. By the Bartlett identities,
\begin{equation*}
\kappa_{jkl}=-(\kappa_{jk,l}+\kappa_{jl,k}+\kappa_{kl,j}+\kappa_{j,k,l}),
\end{equation*}
so no third derivatives are needed.

Let $\hat{\lambda}=\{\hat{\theta}(\tau^2),\hat{\beta}(\tau^2)\}$ solve $U_\theta=U_\beta=0$ at the true $\tau^2$, with elements $\hat{\lambda}_\theta=\hat{\theta}(\tau^2)$ and $\hat{\lambda}_\beta=\hat{\beta}(\tau^2)$. A Taylor expansion of the profile score $U^{\mathrm{PL}}_{\tau^2}(\tau^2)$ in~(\ref{eq.13}) around the true $\lambda$ gives
\begin{equation}
    U^{\mathrm{PL}}_{\tau^2}(\tau^2) = U_{\tau^2} + \sum_{r} U_{\tau^2 r}\,(\hat{\lambda}_r-\lambda_r) + \frac{1}{2}\sum_{r,s} U_{\tau^2 rs}\,(\hat{\lambda}_r-\lambda_r)(\hat{\lambda}_s-\lambda_s) + O_P(N^{-1/2}),
    \label{eq.S2}
\end{equation}
where the terms without a hat are evaluated at the true $\lambda$. The first term on the right-hand side has mean zero, but the other two terms do not. Since $\hat{\lambda}$ is $\sqrt{N}$-consistent, $\hat{\lambda}_r-\lambda_r$ is of order $N^{-1/2}$, whereas its mean is of order $N^{-1}$. The standard expansion gives $\hat{\lambda}_r-\lambda_r=\sum_s \kappa^{r,s}U_s+O_P(N^{-1})$ and $\boldsymbol{E}(\hat{\lambda}_r-\lambda_r)=\delta_r+O(N^{-2})$ with $\delta_r=\sum_{s,t,u} \kappa^{r,s}\kappa^{t,u}(\kappa_{stu}/2+\kappa_{st,u})$~\citep{S-cox1968general}. We take the expectation of each term in~(\ref{eq.S2}).
\begin{enumerate}
\item[(i)] $\boldsymbol{E}\,U_{\tau^2}=0$.
\item[(ii)] By the Bartlett identity, $\boldsymbol{E}\,U_{\tau^2 r}=-\boldsymbol{E}(U_{\tau^2}U_r)=-\kappa_{\tau^2,r}$, which is of order $N$. Write $U_{\tau^2 r}=-\kappa_{\tau^2,r}+\Delta_{r}$, where $\boldsymbol{E}\Delta_{r}=0$ and $\Delta_r$ is of order $N^{1/2}$. The mean of the second term has two sources of order 1: $-\kappa_{\tau^2,r}$ times the mean of $\hat{\lambda}_r-\lambda_r$, and the covariance between $\Delta_r$ and $\hat{\lambda}_r$. That is,
\begin{equation*}
\boldsymbol{E}\{U_{\tau^2 r}(\hat{\lambda}_r-\lambda_r)\}=-\kappa_{\tau^2,r}\delta_r+\sum_s \kappa^{r,s}\boldsymbol{E}(\Delta_{r}U_s)+o(1)=-\kappa_{\tau^2,r}\delta_r+\sum_s \kappa^{r,s}\kappa_{\tau^2 r,s}+o(1).
\end{equation*}
\item[(iii)] Since $U_{\tau^2 rs}=\kappa_{\tau^2 rs}+O_P(N^{1/2})$ is of order $N$ and $\boldsymbol{E}\{(\hat{\lambda}_r-\lambda_r)(\hat{\lambda}_s-\lambda_s)\}=\kappa^{r,s}+O(N^{-2})$, the quadratic term has expectation $\frac{1}{2}\sum_{r,s}\kappa_{\tau^2 rs}\kappa^{r,s}+o(1)$.
\end{enumerate}
Summing (i)--(iii) over $r,s$ gives~(\ref{eq.17}). Each of the three parts of $m(\xi)$ is $O(1)$, and the argument uses expectations rather than stochastic orders only. Since the information for $\tau^2$ is of order $N$, solving the uncentered profile score equation gives $\hat{\tau}^2_{\MLEadj}$ with a bias of order $N^{-1}$.

The three parts of $m(\xi)$ correspond to three sources of the nonzero mean of the profile score. The last part comes from the curvature of $U_{\tau^2}$ in $\lambda$, and the middle part from the covariance between $\hat{\lambda}$ and the derivatives of $U_{\tau^2}$. The first part arises because, under publication bias (PB), $\tau^2$ is not orthogonal to $\lambda$ (that is, $\kappa_{\tau^2,\theta}\ne 0$) and $\hat{\lambda}$ has a bias of order $N^{-1}$. In the normal-normal (NN) model without PB, with $w_i$ and $W$ defined as in Section~\ref{sec.2.1}, $\lambda=\theta$, and the three parts are $0$, $-\sum_i w_i^2/W$, and $\sum_i w_i^2/(2W)$, respectively, as shown in~\ref{ap:B}. Thus, $m=-\sum_i w_i^2/(2W)$, and~(\ref{eq.15}) is the REML equation~(\ref{eq.6}). Under PB, the middle and last parts involve the selection parameter through the conditional density of published studies, and the first part has no counterpart in the REML. Thus, the proposed adjustment is not a plug-in of the REML correction. Centering the profile score by $m(\xi)$ removes its $O(1)$ mean, which is the part of the $O(N^{-1})$ bias of $\hat{\tau}^2_{\MLEadj}$ caused by estimating $\lambda$. As a numerical illustration, for $N=25$, $\tau=0.5$, $\bar{\sigma}=0.3$, $\beta=1$, and $p=0.7$, the three parts of $m(\xi)$ were $+0.28$, $-2.05$, and $+1.30$, respectively.

%======================================================================
\section{Reduction to the restricted maximum likelihood estimator}\label{ap:B}

Without PB, $\ell_{\mathrm{cond},i}$ is the Gaussian log-likelihood, there is no $\beta$, and $\lambda=\theta$. Let $w_i=1/(\tau^2+\sigma_i^2)$ and $W=\sum_i w_i$. Then $U_{\tau^2}=\frac{1}{2}\sum_i\{w_i^2(\hat{\theta}_i-\theta)^2-w_i\}$, $U_\theta=\sum_i w_i(\hat{\theta}_i-\theta)$, and
\begin{equation*}
\begin{aligned}
&\kappa_{\tau^2,\theta}=0,\qquad \kappa_{\theta,\theta}=W,\qquad \kappa_{\tau^2\theta\theta}=\sum_i w_i^2,\\
&\kappa_{\tau^2\theta,\theta}=\boldsymbol{E}\Big\{-\sum_i w_i^2(\hat{\theta}_i-\theta)\sum_{i'} w_{i'}(\hat{\theta}_{i'}-\theta)\Big\}=-\sum_i w_i^2.
\end{aligned}
\end{equation*}
Since $\kappa_{\tau^2,\theta}=0$, the first part of $m(\xi)$ is zero, and
\begin{equation*}
m=-\frac{\sum_i w_i^2}{W}+\frac{1}{2}\frac{\sum_i w_i^2}{W}=-\frac{1}{2}\frac{\sum_i w_i^2}{W}.
\end{equation*}
The estimating equations~(\ref{eq.16}) become
\begin{equation*}
\sum_i w_i(\hat{\theta}_i-\theta)=0,\qquad \frac{1}{2}\left\{\sum_i w_i^2(\hat{\theta}_i-\hat{\theta})^2-\sum_i w_i+\frac{\sum_i w_i^2}{W}\right\}=0,
\end{equation*}
which give $\hat{\theta}=\sum_iw_i\hat{\theta}_i/W$ and
\begin{equation*}
\hat{\tau}^2_{\REMLadj}=\frac{\sum_{i} w_i^2 \left\{(\hat{\theta}_i - \hat{\theta})^2 - \sigma_i^2\right\} }{\sum_{i} w_i^2} + \frac{1}{W},
\end{equation*}
which is identical to the REML estimator~(\ref{eq.5}). Without PB there is no $\beta$ and no penalty, so~(\ref{eq.19}) is also the REML, that is, $\hat{\tau}^2_{\mathrm{Proposed}}=\hat{\tau}^2_{\REMLadj}=\hat{\tau}^2_{\REMLunadj}$.

\textit{Numerical confirmation.} We also confirmed the equivalence numerically with our implementation, in which $m(\xi)$ is computed by the general procedure in~\ref{ap:D} (Gauss--Hermite quadrature, finite-difference derivatives, and the Bartlett identity for the third-order cumulants) without using the closed form above. We fixed $\beta=0$, so that the conditional density of a published study reduces to the normal density, took $\lambda=\theta$, and solved $U_\theta=0$ and $U_{\tau^2}-m=0$. We compared the solutions with the REML estimates obtained by the \pkg{R} package \pkg{metafor}~\citep{S-viechtbauer2010conducting}. For the clopidogrel meta-analysis, the REML estimates of $(\theta,\tau^2)$ were $(-0.54693, 0.09546)$ and our solutions were $(-0.54693, 0.09546)$; for the nicotine gum meta-analysis, they were $(0.51563, 0.08565)$ and $(0.51563, 0.08566)$, respectively. In addition, we generated 1200 meta-analyses without selection, 50 for each combination of $S=15, 25, 50$, $\tau=0.25, 0.5, 0.75, 1$, and $\bar{\sigma}=0.3, 0.5$, under the settings of Section~\ref{sec.4}. The maximum absolute differences from the REML estimates were $2.1\times 10^{-6}$ for $\theta$ and $9.1\times 10^{-6}$ for $\tau^2$, and in all 45 meta-analyses with the REML estimate $\hat{\tau}^2=0$, our solution was also 0. The small differences are due to the finite-difference derivatives.

%======================================================================
\section{The Jeffreys-type penalty on the selection parameter}\label{ap:C}

\textit{Why a penalty is needed.} With a small number of published studies, the conditional log-likelihood $\ell_{\mathrm{cond}}$ in~(\ref{eq.11}) is often monotone in $\beta$, similar to separation in logistic regression, and the maximizer of $\beta$ does not exist. In the original implementation of the MLE, $\beta$ is bounded by $\pm 10/\{\max_i(\hat{\theta}_i)-\min_i(\hat{\theta}_i)\}$. In our simulation studies in Section~\ref{sec.4}, the estimate of $\beta$ by the MLE lay on this bound in 25.5\%--52.6\% of the replicates at $S=15$, 13.5\%--40.8\% at $S=25$, and 3.4\%--20.2\% at $S=50$. When $\hat{\beta}$ lies on an arbitrary bound, the estimates of $\theta$ and $\tau^2$ depend on the bound, and the adjustment $m(\xi)$, which is evaluated at $\hat{\beta}$, can be unstable.

\textit{The penalty.} We use
\begin{equation*}
\mathrm{pen}(\xi)=-\frac{1}{2}\log\big[K(\xi)^{-1}\big]_{\beta\beta},
\end{equation*}
that is, half the logarithm of the efficient information for $\beta$, where $K(\xi)=(\kappa_{j,k})$ is the $3\times 3$ expected information matrix. This is the Jeffreys prior for $\beta$ with $(\theta,\tau^2)$ treated as nuisance parameters, in the spirit of the bias-reducing penalty of~\citet{S-firth1993bias}. As $|\beta|\to\infty$, the selection becomes deterministic in the sign of $\hat{\theta}_i$, the information on $\beta$ vanishes, and $\mathrm{pen}(\xi)\to-\infty$; thus, the penalized equations have a finite solution for $\beta$, similar to the Jeffreys-prior penalty in binomial-response models~\citep{S-kosmidis2021jeffreys}. Unlike a normal or Cauchy prior centered at zero, the penalty does not shrink $\beta$ toward zero: the efficient information for $\beta$ is small near $\beta=0$, where $\theta$ and $\beta$ are nearly confounded. In our simulation studies, the estimate of $\beta$ by our estimator lay on the bound, which is kept only as a safeguard, in at most 2.9\% of the replicates in every setting. Since $\mathrm{pen}(\xi)=O(1)$, it does not change the first-order asymptotic distribution of the estimator.

\textit{Why the adjustment changes.} With the penalty, $\hat{\lambda}=\{\hat{\theta}(\tau^2),\hat{\beta}(\tau^2)\}$ solves $U_r+\partial\,\mathrm{pen}/\partial\lambda_r=0$ for $r\in\{\theta,\beta\}$. A Taylor expansion shows that its first-order bias becomes
\begin{equation*}
\tilde{\delta}_r=\delta_r+\sum_s\kappa^{r,s}\frac{\partial\,\mathrm{pen}(\xi)}{\partial\lambda_s}+O(N^{-2}).
\end{equation*}
The derivation in~\ref{ap:A} goes through with $\delta_r$ replaced by $\tilde{\delta}_r$. Only the first part of $m(\xi)$ is affected, and it becomes $-\sum_r \kappa_{\tau^2,r}\tilde{\delta}_r$, which gives $\tilde{m}(\xi)$ in~(\ref{eq.19}). Thus, $\boldsymbol{E}\{U^{\mathrm{PL}}_{\tau^2}(\tau^2)\}=\tilde{m}(\xi)+o(1)$ for the penalized estimates of $\theta$ and $\beta$.

%======================================================================
\section{Computation and variance estimation}\label{ap:D}

\textit{Expectations.} Each expectation under $g_i$ is a one-dimensional integral, which we compute by $Q=40$-point Gauss--Hermite quadrature with nodes $x_q$ and weights $\omega_q$:
\begin{equation*}
\boldsymbol{E}_i\{h(Y)\}\approx\sum_{q=1}^{Q}\frac{\omega_q}{\sqrt{\pi}}\,h(y_{iq})\,\frac{\Phi(\hat{\alpha}+\beta y_{iq}/\sigma_i)}{\Phi(z_i)},\qquad y_{iq}=\theta+\sqrt{2}\,(\tau^2+\sigma_i^2)^{1/2} x_q.
\end{equation*}
The weights sum to one up to the machine precision. The first and second derivatives of $\ell_{\mathrm{cond},i}$ with respect to $\xi$ are computed by central finite differences with step $10^{-3}\max(1,|\xi_j|)$, with $\hat{\alpha}$ re-solved at every point. The information $\kappa_{j,k}$ and the cumulants $\kappa_{jk,l}$ and $\kappa_{j,k,l}$ are then computed at the quadrature nodes, and $\kappa_{jkl}$ is obtained from the Bartlett identity in~\ref{ap:A}. The penalty is computed from $K(\xi)$, and its gradient by central differences.

\textit{Solving the estimating equations.} The starting value is the penalized MLE, which maximizes $\ell_{\mathrm{cond}}(\xi)+\mathrm{pen}(\xi)$ over $(\theta,\tau^2,\beta)$. We then solve the joint system~(\ref{eq.20}) by a damped Newton method with a numerical Jacobian; the step is halved until the sum of squared estimating functions decreases, and the iteration stops when the step scaled by the expected information is smaller than $10^{-6}$. The constraint $\tau^2\ge 0$ is imposed by projection. We set $\hat{\tau}^2_{\mathrm{Proposed}}=0$ when the adjusted equation for $\tau^2$ is non-positive at $\tau^2=0$, where the equations for $(\theta,\beta)$ are solved with $\tau^2=0$ fixed. The bound for $\beta$ used by the MLE is kept as a safeguard; if the penalized MLE of $\beta$ lies on the bound, $\beta$ is fixed at the bound and removed from $\lambda$.

\textit{Variance estimation.} Let $\Psi_i(\xi)$ be the contribution of the $i$-th study to the estimating functions in~(\ref{eq.20}): the vector of study-level scores $\partial\ell_{\mathrm{cond},i}(\xi)/\partial\xi$ plus $1/N$ times the $O(1)$ adjustment $(\partial_\theta\mathrm{pen},-\tilde{m},\partial_\beta\mathrm{pen})$. We estimate the variance of $\hat{\xi}$ by $\hat{A}^{-1}\hat{B}\hat{A}^{-T}$, where $\hat{A}=\sum_i\partial\Psi_i(\hat{\xi})/\partial\xi^T$ is computed by central differences and $\hat{B}=\sum_i\Psi_i(\hat{\xi})\Psi_i(\hat{\xi})^T$. Parameters on a boundary ($\hat{\tau}^2=0$ or a fixed $\beta$) are excluded. For the MLE, the same estimator is used with $\Psi_i$ equal to the study-level scores. The 95\% confidence interval (CI) for $\theta$ is $\hat{\theta}\pm 1.96\,\widehat{\mathrm{SE}}(\hat{\theta})$, and that for $\tau^2$ is $[\max\{0,\hat{\tau}^2-1.96\,\widehat{\mathrm{SE}}(\hat{\tau}^2)\},\,\hat{\tau}^2+1.96\,\widehat{\mathrm{SE}}(\hat{\tau}^2)]$.

%======================================================================
\section{Complete results of the simulation studies}\label{ap:E}

Tables~\ref{tab:sim_theta_s03}--\ref{tab:sim_tau2_s05} show the complete results of the simulation studies in Section~\ref{sec.4} for $\theta$ and $\tau^2$. We denote the MLE of~\citet{S-copas2013likelihood} by MLE$_{\mathrm{adj}}$, the proposed estimator by Proposed, and the ML and REML without adjustment for PB applied to the published studies and to all generated studies by MLE$_{\mathrm{pub}}$, REML$_{\mathrm{pub}}$, MLE$_{\mathrm{pop}}$, and REML$_{\mathrm{pop}}$, respectively. For the $g$-th replicate ($g=1,\cdots,G$) with the estimate $\hat{\vartheta}_g$ of the true value $\vartheta$, the tables report the convergence rate (Conv., \%); for $\theta$, the median absolute relative bias, $\operatorname{Median}_g\{|\hat{\vartheta}_g-\vartheta|/\vartheta\}$ (MedARB, \%); for $\tau^2$, the median relative bias, $\operatorname{Median}_g\{(\hat{\vartheta}_g-\vartheta)/\vartheta\}$ (MedRB, \%); the mean squared error, $G^{-1}\sum_g(\hat{\vartheta}_g-\vartheta)^2$ (MSE, $\times 10^{-2}$); the empirical coverage probability of the 95\% CI (CP, \%); and the mean length of the 95\% CI (LOCI). All summaries except the convergence rate are computed over the converged replicates. The CIs are the sandwich-based Wald CIs (\ref{ap:D}) for MLE$_{\mathrm{adj}}$ and Proposed, and the Wald CI for $\theta$ and the Q-profile CI for $\tau^2$~\citep{S-viechtbauer2007confidence} for the ML and REML fits.

\clearpage
\begin{sidewaystable}
\centering
\begin{threeparttable}
\caption{Simulation results for $\theta$ (true value 2) under $\bar{\sigma}=0.3$.}
\label{tab:sim_theta_s03}
\footnotesize
\setlength{\tabcolsep}{2.6pt}
\begin{tabular}{ll@{\hspace{14pt}}rrrr@{\hspace{14pt}}rrrr@{\hspace{14pt}}rrrr@{\hspace{14pt}}rrrr}
\hline
 & & \multicolumn{4}{c@{\hspace{14pt}}}{$\tau=0.25$} & \multicolumn{4}{c@{\hspace{14pt}}}{$\tau=0.5$} & \multicolumn{4}{c@{\hspace{14pt}}}{$\tau=0.75$} & \multicolumn{4}{c}{$\tau=1$} \\
\cmidrule(lr){3-6} \cmidrule(lr){7-10} \cmidrule(lr){11-14} \cmidrule(lr){15-18}
Metric & Method & 15 & 25 & 50 & 200 & 15 & 25 & 50 & 200 & 15 & 25 & 50 & 200 & 15 & 25 & 50 & 200 \\
\hline
\multirow{2}{*}{Conv.} & MLE$_{\mathrm{adj}}$ & 97.1 & 96.4 & 97.8 & 97.8 & 98.6 & 99.3 & 99.0 & 99.5 & 99.3 & 99.5 & 98.7 & 99.6 & 99.7 & 99.1 & 99.2 & 99.4 \\
 & Proposed & 99.3 & 100.0 & 99.8 & 100.0 & 97.4 & 98.3 & 99.1 & 100.0 & 97.0 & 97.6 & 98.7 & 100.0 & 96.4 & 96.1 & 98.5 & 100.0 \\
\hline
\addlinespace[6pt]
\multirow{6}{*}{MedARB} & MLE$_{\mathrm{adj}}$ & 4.1 & 3.1 & 2.4 & 1.1 & 5.7 & 4.3 & 3.1 & 1.5 & 7.2 & 4.8 & 3.7 & 1.6 & 7.5 & 6.0 & 3.9 & 1.9 \\
 & Proposed & 3.5 & 2.5 & 2.1 & 1.1 & 4.7 & 3.7 & 2.8 & 1.4 & 6.4 & 4.4 & 3.3 & 1.5 & 7.3 & 5.8 & 3.8 & 1.9 \\
 & MLE$_{\mathrm{pub}}$ & 6.8 & 6.7 & 7.0 & 6.7 & 12.1 & 12.1 & 12.0 & 12.3 & 18.3 & 18.1 & 18.2 & 18.3 & 24.6 & 24.7 & 24.4 & 24.7 \\
 & REML$_{\mathrm{pub}}$ & 6.8 & 6.7 & 7.0 & 6.7 & 12.2 & 12.1 & 12.0 & 12.3 & 18.3 & 18.1 & 18.2 & 18.3 & 24.6 & 24.7 & 24.4 & 24.7 \\
 & MLE$_{\mathrm{pop}}$ & 3.0 & 2.2 & 1.6 & 0.7 & 4.3 & 3.4 & 2.1 & 1.2 & 5.8 & 4.4 & 3.0 & 1.6 & 7.9 & 5.5 & 4.2 & 2.1 \\
 & REML$_{\mathrm{pop}}$ & 3.1 & 2.2 & 1.6 & 0.7 & 4.3 & 3.4 & 2.1 & 1.2 & 5.8 & 4.4 & 3.0 & 1.6 & 7.9 & 5.5 & 4.2 & 2.1 \\
\hline
\addlinespace[6pt]
\multirow{6}{*}{MSE} & MLE$_{\mathrm{adj}}$ & 1.37 & 0.93 & 0.61 & 0.12 & 3.15 & 2.08 & 1.02 & 0.22 & 5.38 & 3.03 & 1.80 & 0.31 & 7.54 & 4.61 & 2.13 & 0.35 \\
 & Proposed & 1.00 & 0.64 & 0.42 & 0.11 & 2.14 & 1.42 & 0.76 & 0.20 & 3.88 & 2.13 & 1.29 & 0.28 & 6.01 & 3.50 & 1.77 & 0.34 \\
 & MLE$_{\mathrm{pub}}$ & 2.75 & 2.24 & 2.13 & 1.86 & 7.44 & 6.77 & 6.30 & 6.13 & 16.38 & 14.81 & 14.22 & 13.58 & 28.57 & 26.51 & 25.45 & 24.52 \\
 & REML$_{\mathrm{pub}}$ & 2.76 & 2.25 & 2.14 & 1.86 & 7.46 & 6.78 & 6.30 & 6.13 & 16.39 & 14.82 & 14.23 & 13.58 & 28.58 & 26.52 & 25.45 & 24.52 \\
 & MLE$_{\mathrm{pop}}$ & 0.78 & 0.44 & 0.23 & 0.05 & 1.54 & 1.03 & 0.44 & 0.12 & 3.02 & 1.85 & 0.89 & 0.23 & 5.35 & 3.06 & 1.55 & 0.39 \\
 & REML$_{\mathrm{pop}}$ & 0.78 & 0.44 & 0.23 & 0.05 & 1.54 & 1.03 & 0.44 & 0.12 & 3.02 & 1.85 & 0.89 & 0.23 & 5.35 & 3.06 & 1.55 & 0.39 \\
\hline
\addlinespace[6pt]
\multirow{6}{*}{CP} & MLE$_{\mathrm{adj}}$ & 71.9 & 79.7 & 84.4 & 94.5 & 68.7 & 75.7 & 86.2 & 95.1 & 70.0 & 77.4 & 85.3 & 94.1 & 73.5 & 76.8 & 85.8 & 93.7 \\
 & Proposed & 88.3 & 88.6 & 89.8 & 94.1 & 86.0 & 88.4 & 90.3 & 95.3 & 86.7 & 87.9 & 91.9 & 93.9 & 85.6 & 88.4 & 91.5 & 94.3 \\
 & MLE$_{\mathrm{pub}}$ & 62.0 & 45.8 & 17.1 & 0.0 & 44.4 & 21.4 & 3.3 & 0.0 & 31.6 & 11.5 & 0.6 & 0.0 & 25.4 & 7.0 & 0.3 & 0.0 \\
 & REML$_{\mathrm{pub}}$ & 63.6 & 46.8 & 17.7 & 0.0 & 47.1 & 22.8 & 3.4 & 0.0 & 33.3 & 12.9 & 0.7 & 0.0 & 27.9 & 7.7 & 0.3 & 0.0 \\
 & MLE$_{\mathrm{pop}}$ & 92.6 & 94.8 & 93.5 & 94.9 & 94.3 & 94.0 & 94.0 & 94.4 & 92.9 & 93.7 & 95.0 & 93.5 & 92.7 & 92.8 & 94.2 & 94.2 \\
 & REML$_{\mathrm{pop}}$ & 93.3 & 94.9 & 93.7 & 94.9 & 95.0 & 94.6 & 94.3 & 94.4 & 94.2 & 94.1 & 95.0 & 93.5 & 93.4 & 93.3 & 94.4 & 94.2 \\
\hline
\addlinespace[6pt]
\multirow{6}{*}{LOCI} & MLE$_{\mathrm{adj}}$ & 0.354 & 0.316 & 0.261 & 0.133 & 0.475 & 0.411 & 0.354 & 0.178 & 0.538 & 0.490 & 0.415 & 0.206 & 0.646 & 0.551 & 0.472 & 0.235 \\
 & Proposed & 0.353 & 0.301 & 0.238 & 0.129 & 0.499 & 0.419 & 0.324 & 0.172 & 0.651 & 0.535 & 0.414 & 0.200 & 0.792 & 0.640 & 0.473 & 0.229 \\
 & MLE$_{\mathrm{pub}}$ & 0.333 & 0.257 & 0.182 & 0.092 & 0.450 & 0.351 & 0.253 & 0.128 & 0.588 & 0.463 & 0.334 & 0.169 & 0.737 & 0.586 & 0.422 & 0.212 \\
 & REML$_{\mathrm{pub}}$ & 0.341 & 0.261 & 0.184 & 0.092 & 0.466 & 0.358 & 0.255 & 0.128 & 0.608 & 0.472 & 0.338 & 0.169 & 0.763 & 0.598 & 0.427 & 0.213 \\
 & MLE$_{\mathrm{pop}}$ & 0.322 & 0.251 & 0.179 & 0.090 & 0.481 & 0.374 & 0.268 & 0.135 & 0.661 & 0.518 & 0.372 & 0.187 & 0.849 & 0.672 & 0.480 & 0.242 \\
 & REML$_{\mathrm{pop}}$ & 0.329 & 0.255 & 0.180 & 0.090 & 0.493 & 0.380 & 0.270 & 0.135 & 0.677 & 0.525 & 0.375 & 0.187 & 0.870 & 0.682 & 0.484 & 0.242 \\
\hline
\end{tabular}
\begin{tablenotes}[flushleft]\footnotesize
\item Note: Conv., convergence rate (\%); MedARB, median absolute relative bias (\%); MedRB, median relative bias (\%); MSE, mean squared error ($\times 10^{-2}$); CP, empirical coverage probability of the 95\% CI (\%); LOCI, mean length of the 95\% CI. MLE$_{\mathrm{adj}}$, the MLE of~\citet{copas2013likelihood}; Proposed, the proposed estimator; MLE$_{\mathrm{pub}}$ and REML$_{\mathrm{pub}}$, ML and REML on the published studies without adjustment for PB; MLE$_{\mathrm{pop}}$ and REML$_{\mathrm{pop}}$, ML and REML on all generated studies. The CIs are sandwich-based Wald CIs for MLE$_{\mathrm{adj}}$ and Proposed, and the Wald CI for $\theta$ and the Q-profile CI for $\tau^2$ for the other methods. The convergence rate is shown only for MLE$_{\mathrm{adj}}$ and Proposed; the other methods converged in all replicates. All other summaries are computed over the converged replicates.
\end{tablenotes}
\end{threeparttable}
\end{sidewaystable}

\begin{sidewaystable}
\centering
\begin{threeparttable}
\caption{Simulation results for $\theta$ (true value 2) under $\bar{\sigma}=0.5$.}
\label{tab:sim_theta_s05}
\footnotesize
\setlength{\tabcolsep}{2.6pt}
\begin{tabular}{ll@{\hspace{14pt}}rrrr@{\hspace{14pt}}rrrr@{\hspace{14pt}}rrrr@{\hspace{14pt}}rrrr}
\hline
 & & \multicolumn{4}{c@{\hspace{14pt}}}{$\tau=0.25$} & \multicolumn{4}{c@{\hspace{14pt}}}{$\tau=0.5$} & \multicolumn{4}{c@{\hspace{14pt}}}{$\tau=0.75$} & \multicolumn{4}{c}{$\tau=1$} \\
\cmidrule(lr){3-6} \cmidrule(lr){7-10} \cmidrule(lr){11-14} \cmidrule(lr){15-18}
Metric & Method & 15 & 25 & 50 & 200 & 15 & 25 & 50 & 200 & 15 & 25 & 50 & 200 & 15 & 25 & 50 & 200 \\
\hline
\multirow{2}{*}{Conv.} & MLE$_{\mathrm{adj}}$ & 98.4 & 98.2 & 98.4 & 99.4 & 98.3 & 99.0 & 98.1 & 98.7 & 99.0 & 99.6 & 98.8 & 99.2 & 99.8 & 99.8 & 99.3 & 99.4 \\
 & Proposed & 99.3 & 99.6 & 99.7 & 100.0 & 98.5 & 98.7 & 99.2 & 99.9 & 98.0 & 98.5 & 98.4 & 99.7 & 97.2 & 96.6 & 97.8 & 99.8 \\
\hline
\addlinespace[6pt]
\multirow{6}{*}{MedARB} & MLE$_{\mathrm{adj}}$ & 5.6 & 4.7 & 3.4 & 1.8 & 7.5 & 6.2 & 4.3 & 2.2 & 8.6 & 6.9 & 5.2 & 2.5 & 9.7 & 7.8 & 5.5 & 2.8 \\
 & Proposed & 4.6 & 3.7 & 3.1 & 1.8 & 6.1 & 4.9 & 3.7 & 2.1 & 7.2 & 5.8 & 4.4 & 2.3 & 8.2 & 6.4 & 4.9 & 2.8 \\
 & MLE$_{\mathrm{pub}}$ & 9.5 & 9.5 & 9.4 & 9.5 & 13.5 & 13.2 & 13.6 & 13.7 & 18.4 & 19.2 & 18.9 & 18.9 & 25.0 & 25.4 & 24.5 & 25.0 \\
 & REML$_{\mathrm{pub}}$ & 9.5 & 9.5 & 9.4 & 9.5 & 13.5 & 13.2 & 13.6 & 13.7 & 18.4 & 19.2 & 18.9 & 18.9 & 25.1 & 25.4 & 24.5 & 25.0 \\
 & MLE$_{\mathrm{pop}}$ & 4.1 & 3.2 & 2.1 & 1.2 & 5.0 & 3.9 & 2.8 & 1.4 & 6.6 & 4.8 & 3.4 & 1.7 & 8.9 & 6.4 & 4.2 & 2.1 \\
 & REML$_{\mathrm{pop}}$ & 4.1 & 3.3 & 2.1 & 1.2 & 5.0 & 3.9 & 2.8 & 1.4 & 6.6 & 4.9 & 3.4 & 1.7 & 8.9 & 6.4 & 4.2 & 2.1 \\
\hline
\addlinespace[6pt]
\multirow{6}{*}{MSE} & MLE$_{\mathrm{adj}}$ & 3.33 & 2.10 & 1.29 & 0.34 & 5.30 & 4.18 & 2.26 & 0.57 & 7.82 & 5.75 & 3.44 & 0.67 & 11.40 & 8.18 & 4.06 & 0.86 \\
 & Proposed & 1.99 & 1.35 & 1.05 & 0.34 & 3.34 & 2.42 & 1.34 & 0.44 & 5.25 & 3.52 & 2.24 & 0.53 & 7.55 & 5.16 & 2.73 & 0.73 \\
 & MLE$_{\mathrm{pub}}$ & 5.42 & 4.63 & 3.96 & 3.73 & 9.71 & 8.73 & 8.09 & 7.56 & 17.21 & 16.81 & 15.55 & 14.55 & 30.52 & 28.83 & 26.39 & 25.44 \\
 & REML$_{\mathrm{pub}}$ & 5.43 & 4.64 & 3.97 & 3.73 & 9.73 & 8.75 & 8.10 & 7.56 & 17.24 & 16.83 & 15.56 & 14.56 & 30.55 & 28.84 & 26.40 & 25.45 \\
 & MLE$_{\mathrm{pop}}$ & 1.50 & 0.93 & 0.46 & 0.11 & 2.30 & 1.41 & 0.65 & 0.17 & 4.00 & 2.24 & 1.10 & 0.28 & 6.18 & 3.75 & 1.75 & 0.42 \\
 & REML$_{\mathrm{pop}}$ & 1.50 & 0.93 & 0.46 & 0.11 & 2.30 & 1.41 & 0.65 & 0.17 & 4.00 & 2.24 & 1.10 & 0.28 & 6.18 & 3.75 & 1.75 & 0.42 \\
\hline
\addlinespace[6pt]
\multirow{6}{*}{CP} & MLE$_{\mathrm{adj}}$ & 77.7 & 82.0 & 86.1 & 92.6 & 72.6 & 73.6 & 81.9 & 93.4 & 73.0 & 77.3 & 81.8 & 94.5 & 75.9 & 75.3 & 85.4 & 94.4 \\
 & Proposed & 89.5 & 89.3 & 87.5 & 92.2 & 85.4 & 86.7 & 89.2 & 94.0 & 84.3 & 88.9 & 90.1 & 95.5 & 87.4 & 85.4 & 91.4 & 94.6 \\
 & MLE$_{\mathrm{pub}}$ & 70.2 & 53.1 & 23.3 & 0.0 & 54.3 & 35.3 & 6.3 & 0.0 & 45.9 & 22.0 & 2.7 & 0.0 & 32.5 & 13.3 & 0.9 & 0.0 \\
 & REML$_{\mathrm{pub}}$ & 71.2 & 53.7 & 23.5 & 0.0 & 56.8 & 36.7 & 6.9 & 0.0 & 47.9 & 23.0 & 2.7 & 0.0 & 35.9 & 14.5 & 0.9 & 0.0 \\
 & MLE$_{\mathrm{pop}}$ & 94.3 & 93.2 & 93.2 & 95.1 & 93.0 & 93.9 & 95.3 & 94.5 & 93.0 & 94.3 & 94.5 & 95.0 & 92.3 & 92.9 & 94.2 & 94.7 \\
 & REML$_{\mathrm{pop}}$ & 94.4 & 93.2 & 93.4 & 95.1 & 93.7 & 94.1 & 95.5 & 94.5 & 93.2 & 94.7 & 95.0 & 95.0 & 93.0 & 93.2 & 94.4 & 94.7 \\
\hline
\addlinespace[6pt]
\multirow{6}{*}{LOCI} & MLE$_{\mathrm{adj}}$ & 0.524 & 0.462 & 0.395 & 0.225 & 0.599 & 0.571 & 0.490 & 0.277 & 0.730 & 0.703 & 0.572 & 0.303 & 0.864 & 0.764 & 0.664 & 0.324 \\
 & Proposed & 0.509 & 0.437 & 0.348 & 0.211 & 0.637 & 0.565 & 0.448 & 0.262 & 0.757 & 0.680 & 0.541 & 0.294 & 0.945 & 0.757 & 0.592 & 0.316 \\
 & MLE$_{\mathrm{pub}}$ & 0.514 & 0.397 & 0.277 & 0.137 & 0.580 & 0.446 & 0.318 & 0.160 & 0.683 & 0.542 & 0.389 & 0.197 & 0.822 & 0.648 & 0.467 & 0.236 \\
 & REML$_{\mathrm{pub}}$ & 0.520 & 0.400 & 0.278 & 0.137 & 0.596 & 0.454 & 0.321 & 0.161 & 0.706 & 0.553 & 0.393 & 0.198 & 0.851 & 0.661 & 0.472 & 0.236 \\
 & MLE$_{\mathrm{pop}}$ & 0.471 & 0.363 & 0.255 & 0.128 & 0.584 & 0.455 & 0.325 & 0.163 & 0.732 & 0.579 & 0.415 & 0.209 & 0.908 & 0.717 & 0.514 & 0.259 \\
 & REML$_{\mathrm{pop}}$ & 0.480 & 0.368 & 0.257 & 0.129 & 0.598 & 0.461 & 0.328 & 0.164 & 0.750 & 0.588 & 0.418 & 0.209 & 0.931 & 0.727 & 0.517 & 0.259 \\
\hline
\end{tabular}
\begin{tablenotes}[flushleft]\footnotesize
\item Note: Conv., convergence rate (\%); MedARB, median absolute relative bias (\%); MedRB, median relative bias (\%); MSE, mean squared error ($\times 10^{-2}$); CP, empirical coverage probability of the 95\% CI (\%); LOCI, mean length of the 95\% CI. MLE$_{\mathrm{adj}}$, the MLE of~\citet{copas2013likelihood}; Proposed, the proposed estimator; MLE$_{\mathrm{pub}}$ and REML$_{\mathrm{pub}}$, ML and REML on the published studies without adjustment for PB; MLE$_{\mathrm{pop}}$ and REML$_{\mathrm{pop}}$, ML and REML on all generated studies. The CIs are sandwich-based Wald CIs for MLE$_{\mathrm{adj}}$ and Proposed, and the Wald CI for $\theta$ and the Q-profile CI for $\tau^2$ for the other methods. The convergence rate is shown only for MLE$_{\mathrm{adj}}$ and Proposed; the other methods converged in all replicates. All other summaries are computed over the converged replicates.
\end{tablenotes}
\end{threeparttable}
\end{sidewaystable}

\begin{sidewaystable}
\centering
\begin{threeparttable}
\caption{Simulation results for $\tau^2$ (true value $\tau^2$) under $\bar{\sigma}=0.3$.}
\label{tab:sim_tau2_s03}
\footnotesize
\setlength{\tabcolsep}{2.6pt}
\begin{tabular}{ll@{\hspace{14pt}}rrrr@{\hspace{14pt}}rrrr@{\hspace{14pt}}rrrr@{\hspace{14pt}}rrrr}
\hline
 & & \multicolumn{4}{c@{\hspace{14pt}}}{$\tau=0.25$} & \multicolumn{4}{c@{\hspace{14pt}}}{$\tau=0.5$} & \multicolumn{4}{c@{\hspace{14pt}}}{$\tau=0.75$} & \multicolumn{4}{c}{$\tau=1$} \\
\cmidrule(lr){3-6} \cmidrule(lr){7-10} \cmidrule(lr){11-14} \cmidrule(lr){15-18}
Metric & Method & 15 & 25 & 50 & 200 & 15 & 25 & 50 & 200 & 15 & 25 & 50 & 200 & 15 & 25 & 50 & 200 \\
\hline
\multirow{2}{*}{Conv.} & MLE$_{\mathrm{adj}}$ & 97.1 & 96.4 & 97.8 & 97.8 & 98.6 & 99.3 & 99.0 & 99.5 & 99.3 & 99.5 & 98.7 & 99.6 & 99.7 & 99.1 & 99.2 & 99.4 \\
 & Proposed & 99.3 & 100.0 & 99.8 & 100.0 & 97.4 & 98.3 & 99.1 & 100.0 & 97.0 & 97.6 & 98.7 & 100.0 & 96.4 & 96.1 & 98.5 & 100.0 \\
\hline
\addlinespace[6pt]
\multirow{6}{*}{MedRB} & MLE$_{\mathrm{adj}}$ & -34.8 & -17.8 & -9.9 & -4.5 & -16.4 & -12.8 & -4.4 & -1.2 & -21.2 & -11.3 & -5.1 & -1.1 & -18.0 & -7.1 & -2.6 & -0.9 \\
 & Proposed & -17.3 & -10.9 & -6.6 & -3.1 & -5.4 & -9.3 & -3.4 & -0.6 & -12.1 & -7.5 & -4.0 & -0.9 & -11.0 & -6.6 & -2.2 & -0.7 \\
 & MLE$_{\mathrm{pub}}$ & -81.6 & -72.0 & -65.8 & -63.2 & -57.3 & -56.4 & -52.4 & -50.4 & -56.7 & -55.5 & -51.6 & -50.0 & -56.8 & -53.3 & -51.0 & -50.6 \\
 & REML$_{\mathrm{pub}}$ & -70.6 & -65.1 & -62.3 & -62.4 & -51.7 & -53.3 & -50.7 & -50.0 & -52.5 & -52.9 & -50.3 & -49.7 & -53.1 & -50.9 & -49.8 & -50.4 \\
 & MLE$_{\mathrm{pop}}$ & -19.5 & -10.4 & -3.6 & -2.7 & -7.6 & -6.0 & -3.0 & -0.8 & -8.7 & -6.1 & -2.6 & -1.2 & -9.7 & -3.7 & -2.6 & -0.8 \\
 & REML$_{\mathrm{pop}}$ & -7.7 & -3.9 & -0.5 & -1.9 & -0.9 & -2.2 & -1.1 & -0.3 & -3.3 & -3.1 & -0.8 & -0.8 & -4.7 & -0.9 & -1.1 & -0.5 \\
\hline
\addlinespace[6pt]
\multirow{6}{*}{MSE} & MLE$_{\mathrm{adj}}$ & 0.50 & 0.33 & 0.20 & 0.05 & 2.65 & 1.80 & 0.94 & 0.26 & 9.75 & 5.77 & 3.34 & 0.82 & 21.24 & 14.42 & 7.47 & 1.96 \\
 & Proposed & 0.57 & 0.33 & 0.19 & 0.05 & 3.10 & 1.90 & 0.90 & 0.25 & 11.94 & 5.84 & 3.19 & 0.81 & 25.46 & 16.10 & 7.62 & 1.97 \\
 & MLE$_{\mathrm{pub}}$ & 0.24 & 0.21 & 0.19 & 0.16 & 2.36 & 2.14 & 1.82 & 1.60 & 11.11 & 9.86 & 8.66 & 8.02 & 33.55 & 29.81 & 26.75 & 25.69 \\
 & REML$_{\mathrm{pub}}$ & 0.22 & 0.19 & 0.18 & 0.16 & 2.09 & 1.96 & 1.72 & 1.58 & 9.98 & 9.11 & 8.26 & 7.91 & 30.13 & 27.65 & 25.62 & 25.40 \\
 & MLE$_{\mathrm{pop}}$ & 0.18 & 0.12 & 0.07 & 0.02 & 1.05 & 0.60 & 0.33 & 0.08 & 4.44 & 2.40 & 1.13 & 0.31 & 11.53 & 6.78 & 3.23 & 0.82 \\
 & REML$_{\mathrm{pop}}$ & 0.20 & 0.12 & 0.07 & 0.02 & 1.13 & 0.62 & 0.34 & 0.08 & 4.77 & 2.47 & 1.16 & 0.31 & 12.10 & 7.02 & 3.29 & 0.82 \\
\hline
\addlinespace[6pt]
\multirow{6}{*}{CP} & MLE$_{\mathrm{adj}}$ & 92.0 & 91.6 & 89.3 & 93.8 & 77.7 & 77.4 & 86.3 & 93.0 & 69.6 & 80.2 & 85.4 & 93.7 & 74.7 & 80.7 & 88.5 & 94.0 \\
 & Proposed & 97.8 & 94.9 & 93.0 & 94.5 & 83.5 & 84.5 & 90.3 & 93.6 & 78.3 & 87.1 & 90.4 & 93.8 & 82.6 & 87.0 & 91.3 & 94.4 \\
 & MLE$_{\mathrm{pub}}$ & 89.4 & 84.5 & 70.5 & 12.7 & 84.3 & 70.0 & 42.7 & 0.4 & 75.3 & 56.3 & 25.7 & 0.0 & 72.2 & 53.3 & 19.4 & 0.0 \\
 & REML$_{\mathrm{pub}}$ & 89.4 & 84.5 & 70.5 & 12.7 & 84.3 & 70.0 & 42.7 & 0.4 & 75.3 & 56.3 & 25.7 & 0.0 & 72.2 & 53.3 & 19.4 & 0.0 \\
 & MLE$_{\mathrm{pop}}$ & 96.4 & 96.6 & 93.8 & 95.9 & 94.4 & 95.8 & 94.5 & 94.2 & 93.6 & 94.5 & 95.4 & 94.3 & 94.2 & 94.3 & 95.7 & 95.8 \\
 & REML$_{\mathrm{pop}}$ & 96.4 & 96.6 & 93.8 & 95.9 & 94.4 & 95.8 & 94.5 & 94.2 & 93.6 & 94.5 & 95.4 & 94.3 & 94.2 & 94.3 & 95.7 & 95.8 \\
\hline
\addlinespace[6pt]
\multirow{6}{*}{LOCI} & MLE$_{\mathrm{adj}}$ & 0.208 & 0.180 & 0.140 & 0.085 & 0.454 & 0.397 & 0.351 & 0.196 & 0.809 & 0.751 & 0.638 & 0.348 & 1.355 & 1.211 & 1.018 & 0.543 \\
 & Proposed & 0.242 & 0.192 & 0.147 & 0.085 & 0.550 & 0.450 & 0.360 & 0.194 & 1.029 & 0.875 & 0.682 & 0.345 & 1.718 & 1.452 & 1.084 & 0.542 \\
 & MLE$_{\mathrm{pub}}$ & 0.181 & 0.121 & 0.079 & 0.041 & 0.418 & 0.277 & 0.180 & 0.085 & 0.729 & 0.488 & 0.318 & 0.148 & 1.147 & 0.786 & 0.510 & 0.235 \\
 & REML$_{\mathrm{pub}}$ & 0.181 & 0.121 & 0.079 & 0.041 & 0.418 & 0.277 & 0.180 & 0.085 & 0.729 & 0.488 & 0.318 & 0.148 & 1.147 & 0.786 & 0.510 & 0.235 \\
 & MLE$_{\mathrm{pop}}$ & 0.216 & 0.155 & 0.105 & 0.050 & 0.512 & 0.356 & 0.238 & 0.114 & 0.972 & 0.682 & 0.458 & 0.217 & 1.598 & 1.147 & 0.762 & 0.363 \\
 & REML$_{\mathrm{pop}}$ & 0.216 & 0.155 & 0.105 & 0.050 & 0.512 & 0.356 & 0.238 & 0.114 & 0.972 & 0.682 & 0.458 & 0.217 & 1.598 & 1.147 & 0.762 & 0.363 \\
\hline
\end{tabular}
\begin{tablenotes}[flushleft]\footnotesize
\item Note: Conv., convergence rate (\%); MedARB, median absolute relative bias (\%); MedRB, median relative bias (\%); MSE, mean squared error ($\times 10^{-2}$); CP, empirical coverage probability of the 95\% CI (\%); LOCI, mean length of the 95\% CI. MLE$_{\mathrm{adj}}$, the MLE of~\citet{copas2013likelihood}; Proposed, the proposed estimator; MLE$_{\mathrm{pub}}$ and REML$_{\mathrm{pub}}$, ML and REML on the published studies without adjustment for PB; MLE$_{\mathrm{pop}}$ and REML$_{\mathrm{pop}}$, ML and REML on all generated studies. The CIs are sandwich-based Wald CIs for MLE$_{\mathrm{adj}}$ and Proposed, and the Wald CI for $\theta$ and the Q-profile CI for $\tau^2$ for the other methods. The convergence rate is shown only for MLE$_{\mathrm{adj}}$ and Proposed; the other methods converged in all replicates. All other summaries are computed over the converged replicates.
\end{tablenotes}
\end{threeparttable}
\end{sidewaystable}

\begin{sidewaystable}
\centering
\begin{threeparttable}
\caption{Simulation results for $\tau^2$ (true value $\tau^2$) under $\bar{\sigma}=0.5$.}
\label{tab:sim_tau2_s05}
\footnotesize
\setlength{\tabcolsep}{2.6pt}
\begin{tabular}{ll@{\hspace{14pt}}rrrr@{\hspace{14pt}}rrrr@{\hspace{14pt}}rrrr@{\hspace{14pt}}rrrr}
\hline
 & & \multicolumn{4}{c@{\hspace{14pt}}}{$\tau=0.25$} & \multicolumn{4}{c@{\hspace{14pt}}}{$\tau=0.5$} & \multicolumn{4}{c@{\hspace{14pt}}}{$\tau=0.75$} & \multicolumn{4}{c}{$\tau=1$} \\
\cmidrule(lr){3-6} \cmidrule(lr){7-10} \cmidrule(lr){11-14} \cmidrule(lr){15-18}
Metric & Method & 15 & 25 & 50 & 200 & 15 & 25 & 50 & 200 & 15 & 25 & 50 & 200 & 15 & 25 & 50 & 200 \\
\hline
\multirow{2}{*}{Conv.} & MLE$_{\mathrm{adj}}$ & 98.4 & 98.2 & 98.4 & 99.4 & 98.3 & 99.0 & 98.1 & 98.7 & 99.0 & 99.6 & 98.8 & 99.2 & 99.8 & 99.8 & 99.3 & 99.4 \\
 & Proposed & 99.3 & 99.6 & 99.7 & 100.0 & 98.5 & 98.7 & 99.2 & 99.9 & 98.0 & 98.5 & 98.4 & 99.7 & 97.2 & 96.6 & 97.8 & 99.8 \\
\hline
\addlinespace[6pt]
\multirow{6}{*}{MedRB} & MLE$_{\mathrm{adj}}$ & -74.5 & -42.5 & -32.4 & -10.0 & -25.2 & -26.8 & -9.7 & -1.3 & -28.2 & -13.6 & -8.7 & -1.4 & -24.0 & -16.0 & -3.1 & -1.8 \\
 & Proposed & -56.4 & -33.3 & -20.0 & -6.1 & -9.5 & -18.6 & -5.1 & -0.2 & -16.9 & -6.8 & -5.5 & -0.5 & -12.0 & -8.7 & -1.7 & -1.1 \\
 & MLE$_{\mathrm{pub}}$ & -100.0 & -100.0 & -100.0 & -100.0 & -74.2 & -72.9 & -67.1 & -63.0 & -64.2 & -58.5 & -56.8 & -54.1 & -60.7 & -57.6 & -53.7 & -52.3 \\
 & REML$_{\mathrm{pub}}$ & -100.0 & -100.0 & -100.0 & -100.0 & -65.6 & -67.7 & -64.5 & -62.3 & -58.6 & -55.0 & -55.0 & -53.6 & -56.1 & -54.7 & -52.2 & -52.0 \\
 & MLE$_{\mathrm{pop}}$ & -42.9 & -25.1 & -17.5 & -3.0 & -13.2 & -9.8 & -4.9 & -1.7 & -13.6 & -7.5 & -2.2 & -0.8 & -11.4 & -5.7 & -3.6 & -1.0 \\
 & REML$_{\mathrm{pop}}$ & -19.4 & -11.6 & -10.5 & -1.3 & -3.2 & -4.0 & -1.9 & -1.0 & -7.0 & -3.5 & -0.2 & -0.3 & -5.6 & -2.2 & -2.0 & -0.5 \\
\hline
\addlinespace[6pt]
\multirow{6}{*}{MSE} & MLE$_{\mathrm{adj}}$ & 1.58 & 1.34 & 0.67 & 0.24 & 5.91 & 3.98 & 2.40 & 0.70 & 14.25 & 9.99 & 6.50 & 1.65 & 35.26 & 22.44 & 13.02 & 3.42 \\
 & Proposed & 1.99 & 1.49 & 0.72 & 0.24 & 7.20 & 4.14 & 2.40 & 0.68 & 17.97 & 11.09 & 6.65 & 1.59 & 42.46 & 25.16 & 13.21 & 3.29 \\
 & MLE$_{\mathrm{pub}}$ & 0.37 & 0.32 & 0.32 & 0.35 & 3.54 & 3.36 & 3.00 & 2.57 & 14.54 & 12.14 & 10.79 & 9.35 & 38.98 & 34.53 & 30.39 & 27.80 \\
 & REML$_{\mathrm{pub}}$ & 0.39 & 0.32 & 0.31 & 0.34 & 3.20 & 3.08 & 2.82 & 2.52 & 13.01 & 11.07 & 10.22 & 9.20 & 34.86 & 31.81 & 28.95 & 27.42 \\
 & MLE$_{\mathrm{pop}}$ & 0.58 & 0.38 & 0.22 & 0.07 & 2.16 & 1.45 & 0.65 & 0.18 & 6.08 & 3.58 & 2.03 & 0.45 & 15.00 & 8.65 & 4.55 & 1.09 \\
 & REML$_{\mathrm{pop}}$ & 0.71 & 0.43 & 0.23 & 0.07 & 2.36 & 1.51 & 0.67 & 0.18 & 6.33 & 3.70 & 2.08 & 0.45 & 15.80 & 8.90 & 4.61 & 1.10 \\
\hline
\addlinespace[6pt]
\multirow{6}{*}{CP} & MLE$_{\mathrm{adj}}$ & 96.5 & 95.2 & 96.9 & 97.6 & 86.2 & 83.2 & 82.6 & 92.1 & 74.9 & 77.7 & 80.9 & 93.7 & 74.3 & 77.3 & 83.9 & 93.5 \\
 & Proposed & 98.2 & 97.2 & 97.6 & 97.9 & 94.1 & 93.8 & 89.7 & 93.1 & 86.6 & 87.3 & 87.2 & 94.4 & 84.8 & 85.7 & 88.3 & 94.4 \\
 & MLE$_{\mathrm{pub}}$ & 91.0 & 84.1 & 70.6 & 17.1 & 88.0 & 77.1 & 55.9 & 3.3 & 80.9 & 70.9 & 41.3 & 0.3 & 77.2 & 60.1 & 28.4 & 0.0 \\
 & REML$_{\mathrm{pub}}$ & 91.0 & 84.1 & 70.6 & 17.1 & 88.0 & 77.1 & 55.9 & 3.3 & 80.9 & 70.9 & 41.3 & 0.3 & 77.2 & 60.1 & 28.4 & 0.0 \\
 & MLE$_{\mathrm{pop}}$ & 95.4 & 95.3 & 94.5 & 92.8 & 96.4 & 94.9 & 96.1 & 94.2 & 96.1 & 95.4 & 94.5 & 95.7 & 94.8 & 95.2 & 94.4 & 94.2 \\
 & REML$_{\mathrm{pop}}$ & 95.4 & 95.3 & 94.5 & 92.8 & 96.4 & 94.9 & 96.1 & 94.2 & 96.1 & 95.4 & 94.5 & 95.7 & 94.8 & 95.2 & 94.4 & 94.2 \\
\hline
\addlinespace[6pt]
\multirow{6}{*}{LOCI} & MLE$_{\mathrm{adj}}$ & 0.458 & 0.406 & 0.294 & 0.165 & 0.670 & 0.574 & 0.490 & 0.314 & 1.029 & 0.957 & 0.834 & 0.501 & 1.597 & 1.440 & 1.286 & 0.704 \\
 & Proposed & 0.552 & 0.452 & 0.303 & 0.165 & 0.869 & 0.690 & 0.523 & 0.314 & 1.295 & 1.109 & 0.893 & 0.500 & 2.060 & 1.676 & 1.323 & 0.704 \\
 & MLE$_{\mathrm{pub}}$ & 0.320 & 0.197 & 0.108 & 0.037 & 0.597 & 0.390 & 0.259 & 0.131 & 0.939 & 0.656 & 0.428 & 0.202 & 1.414 & 0.958 & 0.623 & 0.291 \\
 & REML$_{\mathrm{pub}}$ & 0.320 & 0.197 & 0.108 & 0.037 & 0.597 & 0.390 & 0.259 & 0.131 & 0.939 & 0.656 & 0.428 & 0.202 & 1.414 & 0.958 & 0.623 & 0.291 \\
 & MLE$_{\mathrm{pop}}$ & 0.398 & 0.276 & 0.186 & 0.100 & 0.738 & 0.520 & 0.350 & 0.166 & 1.185 & 0.854 & 0.570 & 0.271 & 1.833 & 1.307 & 0.871 & 0.417 \\
 & REML$_{\mathrm{pop}}$ & 0.398 & 0.276 & 0.186 & 0.100 & 0.738 & 0.520 & 0.350 & 0.166 & 1.185 & 0.854 & 0.570 & 0.271 & 1.833 & 1.307 & 0.871 & 0.417 \\
\hline
\end{tabular}
\begin{tablenotes}[flushleft]\footnotesize
\item Note: Conv., convergence rate (\%); MedARB, median absolute relative bias (\%); MedRB, median relative bias (\%); MSE, mean squared error ($\times 10^{-2}$); CP, empirical coverage probability of the 95\% CI (\%); LOCI, mean length of the 95\% CI. MLE$_{\mathrm{adj}}$, the MLE of~\citet{copas2013likelihood}; Proposed, the proposed estimator; MLE$_{\mathrm{pub}}$ and REML$_{\mathrm{pub}}$, ML and REML on the published studies without adjustment for PB; MLE$_{\mathrm{pop}}$ and REML$_{\mathrm{pop}}$, ML and REML on all generated studies. The CIs are sandwich-based Wald CIs for MLE$_{\mathrm{adj}}$ and Proposed, and the Wald CI for $\theta$ and the Q-profile CI for $\tau^2$ for the other methods. The convergence rate is shown only for MLE$_{\mathrm{adj}}$ and Proposed; the other methods converged in all replicates. All other summaries are computed over the converged replicates.
\end{tablenotes}
\end{threeparttable}
\end{sidewaystable}
\clearpage

%======================================================================
\section{Additional example: the nicotine gum meta-analysis}\label{ap:F}

We additionally analyzed a larger meta-analysis published in the Cochrane Database of Systematic Reviews, conducted by~\citet{S-stead2012nicotine} to evaluate the effect of nicotine gum on smoking cessation with 56 published studies. The data are presented in Table~\ref{tab:data_nico}; the effect size is the lnOR of smoking cessation, with 0.5 added to all cells of studies with a zero cell. Although Begg's rank test, Egger's regression test, and the trim-and-fill method all suggested no PB with $p$-values greater than 0.1, \citet{S-lin2018quantifying} applied a resampling-based test for PB and obtained a $p$-value of 0.005. We analyzed this meta-analysis with our proposed method and the MLE of~\citet{S-copas2013likelihood} under $p=1, 0.9, \cdots, 0.2$, with $\beta$ unrestricted in sign. We show the results in Table~\ref{tab:nico}.

The estimates of the common lnOR, the between-study variance, and the global $I^2$ were close between the two methods under every $p$, which is consistent with the asymptotic equivalence of the two estimators when the number of studies is large. With our method, the significant effect vanished at $p=0.6$ (37 unpublished studies). With the MLE, the 95\% CI for the common lnOR included 0 from $p=0.7$ (24 unpublished studies), but this should be interpreted with caution. At $p=0.9$ and $0.8$, the MLE of $\beta$ lay on its bound ($|\beta|\le 4.37$), and its CIs were computed with $\beta$ fixed at the bound; thus, they did not reflect the uncertainty in $\beta$. At $p=0.7$, the MLE of $\beta$ (2.92) was an interior point, but the profile log-likelihood of $\beta$ was nearly flat: it changed by less than 0.03 for $\beta$ between 2.5 and 4.37. The sandwich standard error of $\hat{\beta}$ was 4.84, and this uncertainty was propagated to the other parameters, which led to the wide CIs for $\theta$, $\tau^2$, and $I^2$ (for example, $[0.0\%, 85.0\%]$ for $I^2$). We confirmed that this estimate is the global maximum of the conditional likelihood from multiple starting values. In contrast, our estimate of $\beta$ was an interior point under all $p$ owing to the Jeffreys-type penalty, and the CIs changed smoothly with $p$. Both methods suggested that the effectiveness of nicotine gum on smoking cessation should be interpreted with caution.

\begin{table}[htb]
    \centering
    \begin{threeparttable}
    \caption{Sensitivity analysis of the nicotine gum meta-analysis ($N=56$) under the marginal selection probability $p$.}
    \label{tab:nico}
    \small
    \begin{tabular}{ccccc}
\hline \multicolumn{5}{c}{MLE based} \\
\hline $p$ & \# of unpublished & $\theta$ & $\tau^2$ & $I^2(\%)$ \\
\hline 1.0 & 0 & 0.515[0.391,0.639] & 0.082[0.018,0.170] & 43.1[14.5,61.1] \\
0.9 & 6 & 0.444[0.336,0.553] & 0.124[0.032,0.215] & 53.3[23.1,66.5] \\
0.8 & 14 & 0.318[0.226,0.410] & 0.203[0.079,0.326] & 65.2[42.2,75.1] \\
0.7 & 24 & 0.209[-0.095,0.514] & 0.259[0.000,0.613] & 70.6[0.0,85.0] \\
0.6 & 37 & 0.101[-0.101,0.304] & 0.306[0.046,0.566] & 73.8[29.7,83.9] \\
0.5 & 56 & -0.033[-0.259,0.193] & 0.368[0.075,0.660] & 77.3[40.9,85.9] \\
0.4 & 84 & -0.200[-0.470,0.070] & 0.444[0.103,0.786] & 80.4[48.7,87.9] \\
0.3 & 131 & -0.417[-0.752,-0.083] & 0.544[0.137,0.951] & 83.4[55.8,89.8] \\
0.2 & 224 & -0.727[-1.157,-0.298] & 0.686[0.186,1.185] & 86.4[63.3,91.6] \\
\hline
\end{tabular}

    \vspace{6pt}

    \begin{tabular}{ccccc}
\hline \multicolumn{5}{c}{Proposed} \\
\hline $p$ & \# of unpublished & $\theta$ & $\tau^2$ & $I^2(\%)$ \\
\hline 1.0 & 0 & 0.516[0.390,0.641] & 0.086[0.018,0.170] & 44.2[14.5,61.1] \\
0.9 & 6 & 0.453[0.341,0.565] & 0.140[0.037,0.242] & 56.3[25.7,69.1] \\
0.8 & 14 & 0.342[0.215,0.470] & 0.198[0.043,0.354] & 64.7[28.3,76.6] \\
0.7 & 24 & 0.242[0.105,0.378] & 0.246[0.061,0.431] & 69.4[35.9,79.9] \\
0.6 & 37 & 0.124[-0.027,0.275] & 0.303[0.080,0.526] & 73.7[42.6,82.9] \\
0.5 & 56 & -0.017[-0.209,0.176] & 0.367[0.099,0.634] & 77.2[47.8,85.4] \\
0.4 & 84 & -0.181[-0.418,0.056] & 0.445[0.126,0.763] & 80.4[53.8,87.6] \\
0.3 & 131 & -0.403[-0.707,-0.099] & 0.544[0.158,0.930] & 83.4[59.4,89.6] \\
0.2 & 224 & -0.714[-1.151,-0.277] & 0.694[0.186,1.202] & 86.5[63.2,91.7] \\
\hline
\end{tabular}
    \begin{tablenotes}[flushleft]\footnotesize
    \item Note: \# of unpublished, estimated number of unpublished studies; $\theta$, common lnOR; $\tau^2$, between-study variance; $I^2$, global $I^2$. Each cell shows the estimate with the 95\% CI in brackets. At $p=1$, the rows are the ML (MLE based) and REML (Proposed) estimates with the Wald CI for $\theta$ and the Q-profile CI for $\tau^2$; at $p<1$, the CIs are sandwich-based Wald CIs.
    \end{tablenotes}
    \end{threeparttable}
\end{table}

\begin{table}[p]
    \centering
    \begin{threeparttable}
    \caption{Data of the nicotine gum meta-analysis.}
    \label{tab:data_nico}
    \footnotesize
    \setlength{\tabcolsep}{4pt}
    \begin{tabular}{cccccc@{\hspace{18pt}}cccccc}
\hline
\multicolumn{6}{c@{\hspace{18pt}}}{Studies 1--28} & \multicolumn{6}{c}{Studies 29--56} \\
\cmidrule(lr){1-6} \cmidrule(lr){7-12}
\multirow{2}{*}{Study} & \multicolumn{2}{c}{Nicotine gum} & \multicolumn{2}{c}{Control} & \multirow{2}{*}{Total} & \multirow{2}{*}{Study} & \multicolumn{2}{c}{Nicotine gum} & \multicolumn{2}{c}{Control} & \multirow{2}{*}{Total} \\
\cmidrule(lr){2-5} \cmidrule(lr){8-11}
& Events & Subjects & Events & Subjects & & & Events & Subjects & Events & Subjects & \\
\hline
1 & 1 & 31 & 2 & 31 & 62 & 29 & 9 & 93 & 13 & 182 & 275 \\
2 & 1 & 13 & 3 & 23 & 36 & 30 & 11 & 21 & 10 & 26 & 47 \\
3 & 12 & 99 & 7 & 52 & 151 & 31 & 13 & 424 & 9 & 412 & 836 \\
4 & 24 & 98 & 28 & 103 & 201 & 32 & 30 & 178 & 22 & 186 & 364 \\
5 & 17 & 200 & 14 & 150 & 350 & 33 & 23 & 180 & 15 & 172 & 352 \\
6 & 15 & 59 & 5 & 19 & 78 & 34 & 9 & 30 & 6 & 30 & 60 \\
7 & 21 & 107 & 21 & 105 & 212 & 35 & 75 & 206 & 50 & 211 & 417 \\
8 & 49 & 211 & 19 & 82 & 293 & 36 & 56 & 99 & 37 & 101 & 200 \\
9 & 39 & 410 & 111 & 1208 & 1618 & 37 & 18 & 41 & 10 & 36 & 77 \\
10 & 51 & 146 & 40 & 127 & 273 & 38 & 23 & 60 & 12 & 53 & 113 \\
11 & 17 & 146 & 15 & 147 & 293 & 39 & 31 & 106 & 16 & 100 & 206 \\
12 & 11 & 100 & 9 & 94 & 194 & 40 & 23 & 210 & 6 & 105 & 315 \\
13 & 5 & 106 & 13 & 319 & 425 & 41 & 81 & 729 & 78 & 1377 & 2106 \\
14 & 129 & 600 & 112 & 617 & 1217 & 42 & 61 & 113 & 28 & 103 & 216 \\
15 & 47 & 481 & 42 & 515 & 996 & 43 & 30 & 71 & 14 & 68 & 139 \\
16 & 11 & 112 & 9 & 111 & 223 & 44 & 37 & 131 & 6 & 46 & 177 \\
17 & 16 & 44 & 6 & 20 & 64 & 45 & 75 & 405 & 17 & 203 & 608 \\
18 & 10 & 101 & 8 & 99 & 200 & 46 & 21 & 68 & 5 & 38 & 106 \\
19 & 110 & 552 & 84 & 522 & 1074 & 47 & 30 & 76 & 13 & 78 & 154 \\
20 & 53 & 378 & 42 & 377 & 755 & 48 & 22 & 58 & 9 & 58 & 116 \\
21 & 40 & 402 & 33 & 420 & 822 & 49 & 13 & 30 & 5 & 30 & 60 \\
22 & 22 & 294 & 37 & 629 & 923 & 50 & 24 & 819 & 9 & 817 & 1636 \\
23 & 23 & 58 & 18 & 58 & 116 & 51 & 28 & 96 & 5 & 49 & 145 \\
24 & 30 & 50 & 23 & 50 & 100 & 52 & 19 & 54 & 7 & 60 & 114 \\
25 & 13 & 54 & 11 & 60 & 114 & 53 & 6 & 73 & 3 & 121 & 194 \\
26 & 5 & 84 & 4 & 89 & 173 & 54 & 8 & 46 & 2 & 40 & 86 \\
27 & 30 & 92 & 22 & 90 & 182 & 55 & 101 & 830 & 24 & 831 & 1661 \\
28 & 29 & 116 & 21 & 113 & 229 & 56 & 24 & 205 & 6 & 222 & 427 \\
\hline
\end{tabular}
    \begin{tablenotes}[flushleft]\footnotesize
    \item Note: Events, number of participants who stopped smoking; Subjects, number of participants in the arm; Total, sum of the subjects in the two arms.
    \end{tablenotes}
    \end{threeparttable}
\end{table}


\begin{thebibliography}{}

\bibitem[Bartlett, 1953]{bartlett1953approximate}
Bartlett, M.~S. (1953).
\newblock Approximate confidence intervals.
\newblock {\em Biometrika}, 40(1--2):12--19.

\bibitem[Begg and Berlin, 1988]{begg1988publication}
Begg, C.~B. and Berlin, J.~A. (1988).
\newblock Publication bias: a problem in interpreting medical data.
\newblock {\em Journal of the Royal Statistical Society: Series A (Statistics
  in Society)}, 151(3):419--445.

\bibitem[Chen et~al., 2013]{chen2013high}
Chen, Y., Zhang, Y., Tang, Y., Huang, X., and Xie, Y. (2013).
\newblock High-maintenance-dose clopidogrel in patients undergoing percutaneous
  coronary intervention: a systematic review and meta-analysis.
\newblock {\em PLoS ONE}, 8(10):e78549.

\bibitem[Clark and Linzer, 2015]{clark2015should}
Clark, T.~S. and Linzer, D.~A. (2015).
\newblock Should {I} use fixed or random effects?
\newblock {\em Political Science Research and Methods}, 3(2):399--408.

\bibitem[Copas and Jackson, 2004]{copas2004bound}
Copas, J. and Jackson, D. (2004).
\newblock A bound for publication bias based on the fraction of unpublished
  studies.
\newblock {\em Biometrics}, 60(1):146--153.

\bibitem[Copas, 2013]{copas2013likelihood}
Copas, J.~B. (2013).
\newblock A likelihood-based sensitivity analysis for publication bias in
  meta-analysis.
\newblock {\em Journal of the Royal Statistical Society: Series C (Applied
  Statistics)}, 62(1):47--66.

\bibitem[Cox and Snell, 1968]{cox1968general}
Cox, D.~R. and Snell, E.~J. (1968).
\newblock A general definition of residuals.
\newblock {\em Journal of the Royal Statistical Society: Series B
  (Methodological)}, 30(2):248--265.

\bibitem[Davey et~al., 2011]{davey2011characteristics}
Davey, J., Turner, R.~M., Clarke, M.~J., and Higgins, J. P.~T. (2011).
\newblock Characteristics of meta-analyses and their component studies in the
  {Cochrane Database of Systematic Reviews}: a cross-sectional, descriptive
  analysis.
\newblock {\em BMC Medical Research Methodology}, 11:160.

\bibitem[DerSimonian and Kacker, 2007]{dersimonian2007random}
DerSimonian, R. and Kacker, R. (2007).
\newblock Random-effects model for meta-analysis of clinical trials: an update.
\newblock {\em Contemporary Clinical Trials}, 28(2):105--114.

\bibitem[DerSimonian and Laird, 1986]{dersimonian1986meta}
DerSimonian, R. and Laird, N. (1986).
\newblock Meta-analysis in clinical trials.
\newblock {\em Controlled Clinical Trials}, 7(3):177--188.

\bibitem[DerSimonian and Laird, 2015]{dersimonian2015meta}
DerSimonian, R. and Laird, N. (2015).
\newblock Meta-analysis in clinical trials revisited.
\newblock {\em Contemporary Clinical Trials}, 45:139--145.

\bibitem[Duval and Tweedie, 2000]{duval2000trim}
Duval, S. and Tweedie, R. (2000).
\newblock Trim and fill: a simple funnel-plot--based method of testing and
  adjusting for publication bias in meta-analysis.
\newblock {\em Biometrics}, 56(2):455--463.

\bibitem[Egger et~al., 1997]{egger1997bias}
Egger, M., Smith, G.~D., Schneider, M., and Minder, C. (1997).
\newblock Bias in meta-analysis detected by a simple, graphical test.
\newblock {\em BMJ}, 315(7109):629--634.

\bibitem[Firth, 1993]{firth1993bias}
Firth, D. (1993).
\newblock Bias reduction of maximum likelihood estimates.
\newblock {\em Biometrika}, 80(1):27--38.

\bibitem[Hardy and Thompson, 1996]{hardy1996likelihood}
Hardy, R.~J. and Thompson, S.~G. (1996).
\newblock A likelihood approach to meta-analysis with random effects.
\newblock {\em Statistics in Medicine}, 15(6):619--629.

\bibitem[Harville, 1974]{harville1974bayesian}
Harville, D.~A. (1974).
\newblock {Bayesian} inference for variance components using only error
  contrasts.
\newblock {\em Biometrika}, 61(2):383--385.

\bibitem[Hedges, 1983]{hedges1983random}
Hedges, L.~V. (1983).
\newblock A random effects model for effect sizes.
\newblock {\em Psychological Bulletin}, 93(2):388.

\bibitem[Higgins and Thompson, 2002]{higgins2002quantifying}
Higgins, J.~P. and Thompson, S.~G. (2002).
\newblock Quantifying heterogeneity in a meta-analysis.
\newblock {\em Statistics in Medicine}, 21(11):1539--1558.

\bibitem[H{\"o}nekopp and Linden, 2022]{honekopp2022heterogeneity}
H{\"o}nekopp, J. and Linden, A.~H. (2022).
\newblock Heterogeneity estimates in a biased world.
\newblock {\em PLoS ONE}, 17(2):e0262809.

\bibitem[Hu et~al., 2024]{hu2024sensitivity}
Hu, T., Zhou, Y., and Hattori, S. (2024).
\newblock Sensitivity analysis for publication bias in meta-analysis of sparse
  data based on exact likelihood.
\newblock {\em Biometrics}, 80(3):ujae092.

\bibitem[Hu et~al., 2026]{hu2024likelihood}
Hu, T., Zhou, Y., Zhou, X.-H., and Hattori, S. (2026).
\newblock Sensitivity analysis for publication bias in diagnostic meta-analysis
  of sparsity using the {Copas} $t$-statistic selection function.
\newblock {\em Statistics in Medicine}, 45(6--7):e70465.

\bibitem[Huang et~al., 2021]{huang2021using}
Huang, A., Komukai, S., Friede, T., and Hattori, S. (2021).
\newblock Using clinical trial registries to inform {Copas} selection model for
  publication bias in meta-analysis.
\newblock {\em Research Synthesis Methods}, 12(5):658--673.

\bibitem[Huang et~al., 2023]{huang2023adjusting}
Huang, A., Morikawa, K., Friede, T., and Hattori, S. (2023).
\newblock Adjusting for publication bias in meta-analysis via inverse
  probability weighting using clinical trial registries.
\newblock {\em Biometrics}, 79(3):2089--2102.

\bibitem[Huedo-Medina et~al., 2006]{huedo2006assessing}
Huedo-Medina, T.~B., S{\'a}nchez-Meca, J., Mar{\'\i}n-Mart{\'\i}nez, F., and
  Botella, J. (2006).
\newblock Assessing heterogeneity in meta-analysis: {Q} statistic or {I}$^2$
  index?
\newblock {\em Psychological Methods}, 11(2):193.

\bibitem[Jin et~al., 2015]{jin2015statistical}
Jin, Z.-C., Zhou, X.-H., and He, J. (2015).
\newblock Statistical methods for dealing with publication bias in
  meta-analysis.
\newblock {\em Statistics in Medicine}, 34(2):343--360.

\bibitem[Kauermann and Carroll, 2001]{kauermann2001note}
Kauermann, G. and Carroll, R.~J. (2001).
\newblock A note on the efficiency of sandwich covariance matrix estimation.
\newblock {\em Journal of the American Statistical Association},
  96(456):1387--1396.

\bibitem[Knapp et~al., 2006]{knapp2006assessing}
Knapp, G., Biggerstaff, B.~J., and Hartung, J. (2006).
\newblock Assessing the amount of heterogeneity in random-effects
  meta-analysis.
\newblock {\em Biometrical Journal}, 48(2):271--285.

\bibitem[Kosmidis and Firth, 2021]{kosmidis2021jeffreys}
Kosmidis, I. and Firth, D. (2021).
\newblock {Jeffreys}-prior penalty, finiteness and shrinkage in
  binomial-response generalized linear models.
\newblock {\em Biometrika}, 108(1):71--82.

\bibitem[Kuss, 2015]{kuss2015statistical}
Kuss, O. (2015).
\newblock Statistical methods for meta-analyses including information from
  studies without any events—add nothing to nothing and succeed nevertheless.
\newblock {\em Statistics in Medicine}, 34(7):1097--1116.

\bibitem[Langan et~al., 2019]{langan2019comparison}
Langan, D., Higgins, J.~P., Jackson, D., Bowden, J., Veroniki, A.~A.,
  Kontopantelis, E., Viechtbauer, W., and Simmonds, M. (2019).
\newblock A comparison of heterogeneity variance estimators in simulated
  random-effects meta-analyses.
\newblock {\em Research Synthesis Methods}, 10(1):83--98.

\bibitem[Langan et~al., 2015]{langan2015empirical}
Langan, D., Higgins, J.~P., and Simmonds, M. (2015).
\newblock An empirical comparison of heterogeneity variance estimators in 12
  894 meta-analyses.
\newblock {\em Research Synthesis Methods}, 6(2):195--205.

\bibitem[Lipsey and Wilson, 2017]{lipsey2017efficacy}
Lipsey, M.~W. and Wilson, D.~B. (2017).
\newblock The efficacy of psychological, educational, and behavioral treatment:
  confirmation from meta-analysis.
\newblock In {\em Quantitative methods in criminology}, pages 65--93.
  Routledge.

\bibitem[Mazei et~al., 2015]{mazei2015meta}
Mazei, J., H{\"u}ffmeier, J., Freund, P.~A., Stuhlmacher, A.~F., Bilke, L., and
  Hertel, G. (2015).
\newblock A meta-analysis on gender differences in negotiation outcomes and
  their moderators.
\newblock {\em Psychological Bulletin}, 141(1):85.

\bibitem[McCullagh, 1987]{mccullagh1987tensor}
McCullagh, P. (1987).
\newblock {\em Tensor methods in statistics}.
\newblock Chapman and Hall, London.

\bibitem[McCullagh and Tibshirani, 1990]{mccullagh1990simple}
McCullagh, P. and Tibshirani, R. (1990).
\newblock A simple method for the adjustment of profile likelihoods.
\newblock {\em Journal of the Royal Statistical Society: Series B
  (Methodological)}, 52(2):325--344.

\bibitem[Morris et~al., 2019]{morris2019using}
Morris, T.~P., White, I.~R., and Crowther, M.~J. (2019).
\newblock Using simulation studies to evaluate statistical methods.
\newblock {\em Statistics in Medicine}, 38(11):2074--2102.

\bibitem[Patterson and Thompson, 1971]{patterson1971recovery}
Patterson, H.~D. and Thompson, R. (1971).
\newblock Recovery of inter-block information when block sizes are unequal.
\newblock {\em Biometrika}, 58(3):545--554.

\bibitem[Schmidt et~al., 2009]{schmidt2009fixed}
Schmidt, F.~L., Oh, I.-S., and Hayes, T.~L. (2009).
\newblock Fixed-versus random-effects models in meta-analysis: model properties
  and an empirical comparison of differences in results.
\newblock {\em British Journal of Mathematical and Statistical Psychology},
  62(1):97--128.

\bibitem[Stead et~al., 2012]{stead2012nicotine}
Stead, L.~F., Perera, R., Bullen, C., Mant, D., Hartmann-Boyce, J., Cahill, K.,
  and Lancaster, T. (2012).
\newblock Nicotine replacement therapy for smoking cessation.
\newblock {\em Cochrane Database of Systematic Reviews}, (11).

\bibitem[Stijnen et~al., 2010]{stijnen2010random}
Stijnen, T., Hamza, T.~H., and {\"O}zdemir, P. (2010).
\newblock Random effects meta-analysis of event outcome in the framework of the
  generalized linear mixed model with applications in sparse data.
\newblock {\em Statistics in Medicine}, 29(29):3046--3067.

\bibitem[Viechtbauer, 2005]{viechtbauer2005bias}
Viechtbauer, W. (2005).
\newblock Bias and efficiency of meta-analytic variance estimators in the
  random-effects model.
\newblock {\em Journal of Educational and Behavioral Statistics},
  30(3):261--293.

\bibitem[Viechtbauer, 2007]{viechtbauer2007confidence}
Viechtbauer, W. (2007).
\newblock Confidence intervals for the amount of heterogeneity in
  meta-analysis.
\newblock {\em Statistics in Medicine}, 26(1):37--52.

\bibitem[Viechtbauer, 2010]{viechtbauer2010conducting}
Viechtbauer, W. (2010).
\newblock Conducting meta-analyses in {R} with the metafor package.
\newblock {\em Journal of Statistical Software}, 36(3):1--48.

\bibitem[Zhou et~al., 2023]{zhou2023likelihood}
Zhou, Y., Huang, A., and Hattori, S. (2023).
\newblock A likelihood-based sensitivity analysis for publication bias on the
  summary receiver operating characteristic in meta-analysis of diagnostic test
  accuracy.
\newblock {\em Statistics in Medicine}, 42(6):781--798.

\bibitem[Zhou et~al., 2024]{zhou2024nonparametric}
Zhou, Y., Huang, A., and Hattori, S. (2024).
\newblock Nonparametric worst-case bounds for publication bias on the summary
  receiver operating characteristic curve.
\newblock {\em Biometrics}, 80(3):ujae080.

\end{thebibliography}

\clearpage
\begin{thebibliography}{}

\bibitem[Bartlett, 1953]{S-bartlett1953approximate}
Bartlett, M.~S. (1953).
\newblock Approximate confidence intervals.
\newblock {\em Biometrika}, 40(1--2):12--19.

\bibitem[Copas, 2013]{S-copas2013likelihood}
Copas, J.~B. (2013).
\newblock A likelihood-based sensitivity analysis for publication bias in
  meta-analysis.
\newblock {\em Journal of the Royal Statistical Society: Series C (Applied
  Statistics)}, 62(1):47--66.

\bibitem[Cox and Snell, 1968]{S-cox1968general}
Cox, D.~R. and Snell, E.~J. (1968).
\newblock A general definition of residuals.
\newblock {\em Journal of the Royal Statistical Society: Series B
  (Methodological)}, 30(2):248--265.

\bibitem[Firth, 1993]{S-firth1993bias}
Firth, D. (1993).
\newblock Bias reduction of maximum likelihood estimates.
\newblock {\em Biometrika}, 80(1):27--38.

\bibitem[Kosmidis and Firth, 2021]{S-kosmidis2021jeffreys}
Kosmidis, I. and Firth, D. (2021).
\newblock {Jeffreys}-prior penalty, finiteness and shrinkage in
  binomial-response generalized linear models.
\newblock {\em Biometrika}, 108(1):71--82.

\bibitem[Lin and Chu, 2018]{S-lin2018quantifying}
Lin, L. and Chu, H. (2018).
\newblock Quantifying publication bias in meta-analysis.
\newblock {\em Biometrics}, 74(3):785--794.

\bibitem[McCullagh, 1987]{S-mccullagh1987tensor}
McCullagh, P. (1987).
\newblock {\em Tensor methods in statistics}.
\newblock Chapman and Hall, London.

\bibitem[Stead et~al., 2012]{S-stead2012nicotine}
Stead, L.~F., Perera, R., Bullen, C., Mant, D., Hartmann-Boyce, J., Cahill, K.,
  and Lancaster, T. (2012).
\newblock Nicotine replacement therapy for smoking cessation.
\newblock {\em Cochrane Database of Systematic Reviews}, (11).

\bibitem[Viechtbauer, 2007]{S-viechtbauer2007confidence}
Viechtbauer, W. (2007).
\newblock Confidence intervals for the amount of heterogeneity in
  meta-analysis.
\newblock {\em Statistics in Medicine}, 26(1):37--52.

\bibitem[Viechtbauer, 2010]{S-viechtbauer2010conducting}
Viechtbauer, W. (2010).
\newblock Conducting meta-analyses in {R} with the metafor package.
\newblock {\em Journal of Statistical Software}, 36(3):1--48.

\end{thebibliography}
\end{document}